\documentclass[printer]{aa}  

\bibpunct{(}{)}{;}{a}{}{,} 
\usepackage{ulem}
\usepackage{natbib}
\usepackage{graphicx}
\usepackage{lscape}
\usepackage{rotating}
\usepackage{subfig}
\usepackage{amsmath}
\usepackage{multirow}
\usepackage{adjustbox}
\usepackage{xcolor,colortbl,tabularx}
\usepackage{graphicx}
\usepackage{caption}
\usepackage{booktabs}
\usepackage[draft]{hyperref}
\usepackage[T1]{fontenc}
\usepackage{txfonts}
\newcommand{\nustar}{{\it NuSTAR}\xspace}
\newcommand{\nicer}{{\it NICER}\xspace}
\newcommand{\swift}{{\it Swift}\xspace}
\newcommand{\maxi}{{\it MAXI}\xspace}

\newcommand{\mysou}{MAXI J1820\xspace}
\begin{document}

   \title{Tracking the evolution of the accretion flow in MAXI J1820+070 during its hard state with the JED-SAD model}

  \author{A. Marino\inst{1,2,3}, S. Barnier\inst{4}, P. O. Petrucci\inst{4}, M. Del Santo\inst{2}, J. Malzac\inst{3}, J. Ferreira\inst{4}, G. Marcel\inst{5}, A. Segreto\inst{2}, S. E. Motta\inst{6}, A. D'A\`i\inst{2}, T. Di Salvo\inst{1}, S. Guillot\inst{7,8}, T. D. Russell\inst{2}}

   \institute{Universit\`a degli Studi di
  Palermo, Dipartimento di Fisica e Chimica, via Archirafi 36 - 90123 Palermo, Italy
              \and
             INAF/IASF Palermo, via Ugo La Malfa 153, I-90146 - Palermo, Italy 
             \and 
             IRAP, Universit\`e de Toulouse, CNRS, UPS, CNES, Toulouse, France
             \and
             Univ. Grenoble Alpes, CNRS, IPAG, F-38000 Grenoble, France
             \and
             Institute of Astronomy, University of Cambridge, Madingley Road, Cambridge CB3 0HA, UK
             \and
             Istituto Nazionale di Astrofisica, Osservatorio Astronomico di Brera, via E. Bianchi 46, 23807 Merate (LC), Italy
             \and
             CNRS, IRAP, 9 avenue du Colonel Roche, BP 44346, F-31028 Toulouse Cedex 4, France
             \and
             Universit\`e de Toulouse, CNES, UPS-OMP, F-31028 Toulouse, France
                    }

   \date{Received XXX; accepted XXX}
 
  \abstract{X-ray binaries in outburst typically show two canonical X-ray spectral states, i.e. hard and soft states, as well as different intermediate states, in which the physical properties of the accretion flow are known to change. However, the truncation of the optically thick disk and the geometry of the optically thin accretion flow (corona) in hard state are still debated. Recently, the JED-SAD paradigm has been proposed for black hole X-ray binaries, aimed to address the topic of accretion and ejection, and their interplay in these systems. According to this model, the accretion flow is composed by an outer standard Shakura-Sunyaev disk (SAD) and an inner hot Jet Emitting Disk (JED). The JED produces both the hard X-ray emission, effectively playing the role of the hot corona, and the radio jets. The disruption of the JED at the transition to the soft state coincides with the quenching of the jet.}
  {In this paper, we use the JED-SAD model to describe the evolution of the accretion flow in the black hole transient MAXI~J1820+070 during its hard and hard-intermediate states. Contrarily to the previous applications of this model, the Compton reflection component has been taken into account.}
  {We use eight broadband X-rays spectra, including \nustar, \nicer\ and the {\it{Neil Gehrels Swift Observatory}} data, providing a total spectral coverage of 0.8-190 keV. The data were directly fitted with the JED-SAD model. We performed the procedure twice, considering two different values for the ISCO, i.e. 4 R$_{\rm G}$ (a$_*$=0.55) and 2 R$_{\rm G}$ (a$_*$=0.95).  
  }
  {Our results suggest that the optically thick disk (i.e. the SAD) does not extend down to the ISCO in any of the considered epochs. In particular, assuming $R_{\rm ISCO}$=4 R$_{\rm G}$, as the system evolves towards the hard/intermediate state, we find an inner radius within a range of $\sim$60 R$_{\rm G}$ in the first observation down to $\sim$30 R$_{\rm G}$ in the last one. The decrease of the inner edge of the SAD is accompanied by an increase of the mass-accretion rate. However, when we assume $R_{\rm ISCO}$=2 we find that the mass accretion rate remains constant and the evolution of the accretion flow is driven by the decrease of the sonic Mach number $m_{\rm S}$, which is unexpected. In all hard-intermediate state observations, two reflection components, characterized by different values of ionization, are required to adequately explain the data. These components likely originate from different regions of the SAD. 
  }
  {The analysis performed provides a coherent physical evolution of the accretion flow in the hard and hard-intermediate states and supports a truncated disk scenario. We show that a flared outer disk could, in principle, explain the double reflection component. The odd results obtained for $R_{\rm ISCO}$=2 R$_{\rm G}$ can also be considered as a further evidence that MAXI~J1820+070 harbours a moderately spinning black hole, as suggested in other works.}
 
   \keywords{accretion -- accretion disks -- ISM: jets and outflows -- X-rays: binaries -- X-rays, individuals: MAXI J1820+070
               }

  \authorrunning{Marino, A.}
  \maketitle
  \email{alessio.marino@unipa.it}

\section{Introduction}
A black hole X-ray binary (XRB) is a binary system composed of a stellar mass BH accreting matter from a companion star \citep[see][for a general review]{Done2007}. Almost all known black hole XRBs are transients (BHTs): they spend most of the time in a quiescent state, at low X-ray luminosity (i.e., below $\sim$ 10$^{31}$ erg s$^{-1}$), but can display sudden episodes of increased X-ray activity called outbursts (with X-ray luminosity up to $\sim$ 10$^{38}$-10$^{39}$ erg s$^{-1}$ at the peak). When in outburst, these systems can be found in a variety of accretion states characterized by different broadband X-ray spectral shape and timing properties \citep[see][for reviews]{Remillard2006,Dunn2010,Belloni2016}. In particular, we distinguish between the hard (HS) and soft (SS) spectral states. The HS is characterized by a cut-off (around 100 keV when detected) power-law-like spectrum extending up to high energies, and a cool (kT$_{\rm disc}\sim0.1-0.3 \ {\rm keV}$) multi-color disk blackbody. In the SS, spectra are dominated by a hotter multi-color disk black-body component (kT$_{\rm disc}\sim$0.8-1.0 keV), and sometimes an additional steeper hard tail is detected as well. In all these states, a further contribution to the overall spectrum is given by the reflection component, a component thought to be due to the hot photons emitted by the corona reprocessing on the accretion disk (see below). The study of the reflection component, and in particular of the relativistically broadened Fe K line at 6.4 keV, can be used as a powerful diagnostics tool to infer, e.g., the inner radius and ionization of the accretion disk and the inclination of the system \citep[see e.g.][for a review]{Reynolds2003}. \\ The variety of spectral states in BHTs has traditionally been interpreted as being due to different physical properties and geometry of the accretion flow around the BH. In the HS, it is thought that the emission is due to thermal Comptonisation by a hot, optically thin electron plasma (corona). The source of soft seed photons is the (possibly truncated) optically thick accretion disk \citep{Shakura1973}. Conversely, in the SS such a disk is expected to extend down to the innermost stable circular orbit \citep[ISCO, ][]{Bardeen1972} and only a marginal contribution from the corona is observed. However, the {\it{consensus}} on the truncated disk model is not global and a number of counter-arguments have been given as well \citep[see, e.g. ][ for an extensive discussion on the topic]{Zdziarski2020}. Among the critical issues, spectral modelling of seemingly broad Fe K lines in BHTs indicated that the optically thick disk was close or even coincident with the ISCO well within the hard state \citep[examples of these can be found in][]{Miller2006, Tomsick2008, Garcia2015}. \\
The controversies on the accretion flow geometry across spectral states is one of the open problems regarding BHTs. When transients rise from quiescence, their spectral evolution usually follows a standard path: the system moves from quiescence to the hard state in a rise of luminosity of several orders of magnitude, then evolves through the hard/intermediate (HIMS) and soft/intermediate (SIMS) up to the soft state. Thereafter, the luminosity decreases and the source exhibits the same transition in reverse but at a much lower luminosity. The common behaviour of BHTs in making transitions from hard-to-soft and from soft-to-hard states at different luminosity is called hysteresis \citep{Miyamoto1995,Zdziarski2004}. The origin of such a pattern is a matter of strong debate. 
It is also noteworthy that a relevant fraction of BH binaries \citep[about 40\%, ][]{Tetarenko2016} are "hard-only" or have displayed "failed transition" outbursts \citep[see, e.g. ][]{Hynes2000, Brocksopp2004, Capitanio2009, Ferrigno2012, DelSanto2016, Bassi2019,2021MNRAS.502..521D}. BHTs are known to launch mildly relativistic jets, which account for the emission from these systems over radio and mid-IR frequencies. Jets are observed only in hard and hard intermediate states, while they are "quenched" around the transition to the soft state \citep{Fender1999,Corbel2000, Fender2004,Russell2020} and a short-lived transient jet can be launched \citep[e.g.,][]{Fender2004,Fender2010,2017MNRAS.468.2788R,2019ApJ...883..198R}. A third critical ingredient is represented by how the X-ray timing properties, i.e. X-rays periodic and aperiodic variability, change according to the accretion state of BHTs. In this context, one of the most intriguing challenges consists in the interpretation of Quasi-Periodic Oscillations (QPOs), i.e., peaks observed in the X-rays Fourier Power Density Spectra of BHTs which are correlated with the spectral evolution of the systems \citep[see][for recent reviews]{Ingram2020, Motta2021}. 

\subsection{The JED-SAD paradigm}\label{ss:jed-sad}
Much effort has been dedicated to interpreting the multi-wavelength spectral behaviour and the timing properties of BH binaries in outburst. However, taking into account the correlation between accretion and ejection has proven to be difficult. A first attempt to describe globally such complex behavior was addressed by \cite{Esin1997}, who 
proposed a multi-flow configuration for the disk. According to these authors, the accretion flow around BHs would consist of an outer Shakura-Sunyaev accretion disk and an inner, less dense and radiatively inefficient plasma phase, dubbed Advection Dominated Accretion Flow \citep[ADAF, ][]{Narayan1995, Yuan2001}. However, the model suffered several weaknesses, in particular it was not able to reproduce hard states at luminosities comparable to the Eddington Limit \citep[see the Introduction of ][ for an extensive discussion on the topic]{Marcel2018a}. In the following decade, many updates were made to the model \citep[see, e.g. ][]{Yuan2001,Meier2005,Xie2012} to make it more suitable to describe the increasing number of X-ray observations of BHTs. However, in none of these updated versions of the model the correlation between accretion flow and jets has been addressed. \\ An attempt to connect accretion and ejection in a unified model has been recently reported in a number of papers \citep{Ferreira2006, Marcel2018a, Marcel2018b, Marcel2019, Marcel2020}. This model connects the spectral evolution BHTs in outbursts to changes in the multi-flow configuration of the accretion disk, similarly to the approach of e.g., \cite{Esin1997}. The main novelty of this paradigm consists in its ability to also explain the appearance and disappearance of the jet and their correlation with the accretion flow. According to this model, the accretion disk is threaded with a large scale vertical magnetic field $B_{\rm Z} (R)$. Recent numerical simulations have shown that such magnetic fields become stronger near the compact object \cite[e.g., ][]{Liska2018}. As a consequence, the accretion flow is also expected to become more magnetized as we approach the inner edge of the disk. In the following we define the magnetization as $\mu(R)=\displaystyle\frac{B^2_Z(R)}{\mu_0 P_{\rm tot}}$, with $P_{\rm tot}$ the total (radiation plus gas) pressure. In the outer regions of the disk, $\mu \ll 1$ and particles are barely affected by the presence of the vertical magnetic field. As a consequence, jet launching is inactive in this region of the disk. Indeed, in order to launch magneto-centrifugally driven jets \citep{BlandfordPayne1982}, it has been shown that $\mu \gtrsim$0.1 must be achieved \citep{Ferreira1995,Ferreira1997,JacqueminIde2019}. For $\mu<10^{-3}$ or less, magnetic winds can be launched, possibly carrying away a significant fraction of the mass. However, these winds exert a negligible torque on the underlying disk \citep{Zhu2018,JacqueminIde2021}.  At these distances from the BH, particles are only subject to the torque due to the internal turbulent viscosity, so that the disk can be well described with the classic \cite{Shakura1973} model. The inner regions are instead highly magnetized, with $\mu$ beyond the 0.1 threshold and powerful jets can be launched. Jets carry away mass, energy and angular momentum and exert then an additional torque \citep[see, e.g.][]{Ferreira1993,Ferreira1995}. Subsequently, accretion proceeds here at a supersonic velocity, i.e. much higher than in the Shakura-Sunyaev accretion disk, and plasma in the internal regions of the accretion flow result more rarefied. Summing up, the accretion flow surrounding the BH has a two-flow configuration: it is composed of a Shakura-Sunyaev disk in the outer regions of the flow, defined as Standard Accretion Disk or SAD, and an inner, less dense and optically thin Jet Emitting Disk (JED) \citep{Ferreira2006}. JED shares the same physical properties of the hot corona, but additionally it also drives jets. Quite intuitively, each accretion state could be obtained by mixing these ingredients with different quantities, i.e. with a hybrid JED-SAD configuration where the two realms extend over regions of different scale. In this sense, such configuration is fundamentally determined by two main control parameters: (1) the transition radius $R_{\rm J}$ between the JED and the SAD, and (2) the inner mass accretion flow $\dot{M}_{\rm in}$ feeding the BH. A detailed description of the model has been presented in \cite{Marcel2018a,Marcel2018b} and we refer the reader to these papers for more details. This spectral model has been successfully used to reproduce the X-ray spectral evolution of the BHT GX~339$-$4 during four outbursts between 2001 and 2011 \citep{Marcel2019, Marcel2020}. Moreover, using the $R_{\rm J}-\dot{M}_{\rm in}$ pairs that best described the X-ray spectra in a hard state, these authors were also able to reproduce the jet emission observed simultaneously in the radio band. Finally, it has been recently shown that also some timing features could be explained within the JED-SAD paradigm framework. Indeed, a direct proportionality has been observed between the Keplerian frequency of the transition radius $R_{\rm J}$ and the type-C QPO frequency in four different outbursts of GX~339$-$4 \citep{Marcel2020}. According to these results, this type of QPOs could originate at the interface between two regions of different values of $\mu$, i.e. being then strictly related to the existence of two different types of accretion flow in the hard and hard/intermediate states of BHTs (Ferreira et al. 2021, submitted). \\
The JED-SAD paradigm has successfully explained much of the observed accretion and ejection phenomenology in
GX~339$-$4. However, the variety of different behaviours observed in three decades of BHT studies \citep[see, e.g. ][for observational reviews]{Dunn2010,Tetarenko2016review} demands for other tests of this model. In this paper, we report on the application of the {\it JED}-{\it SAD} model to the BHT MAXI J1820+070 (ASASSN-18ey), hereafter \mysou, exploiting a broad X-ray data set including data from \nustar, the Neil Gehrels {\it Swift Observatory} (hereafter \swift) and the {\it Neutron Star Interior Composition Explorer} (\nicer). This is the first time the JED-SAD model is applied directly to the data through spectral fits and that Compton reflection process is taken into account. It therefore represents a major test for the potentialities of such paradigm.

\subsection{MAXI J1820+070}
\mysou\ is a BHT that was observed for the first time in the optical band by the All-Sky Automated Search for SuperNovae ASSAS-SN \citep{Shappee2014} on 2018 March 3, and one week later by \maxi\ in X-rays \citep{Kawamuro2018}. Detailed studies of the optical counterpart revealed that the system hosts a stellar-mass black hole ($\sim$ 8.5 M$_\odot$) accreting from a $\sim$ 0.4 M$_\odot$ companion star \citep{Torres2019,Torres2020}. Since its discovery, the source underwent a long (approximately one year) and bright outburst, becoming, at its peak, the second brightest object in the X-ray sky. Due to its brightness, the system was the object of an impressive multi-wavelength observing campaign \citep[see, e.g. ][]{Shidatsu2018,Paice2019,Hoang2019, Trushkin2018,Bright2020,Tetarenko2021} and of a large number of studies. The most recent and precise measure of the distance, determined via radio parallax, is 3.0$\pm$0.3 kpc \citep{Atri2020}. Furthermore, the system shows X-ray dips \citep{Kajava2019} but not eclipses, suggesting an inclination between 60$^\circ$ and 80$^\circ$. Other evidences of the high inclination of the system are provided by optical spectroscopy \citep{Torres2019} and by the estimate of the inclination of the jet axis \citep{Atri2020,2021MNRAS.tmp.1471W}, i.e. about 63$^\circ$. The orbital period of the system has also been estimated being around 0.68 days \citep{Patterson2018, Torres2020}. In X-rays, the outburst was studied in detail in hard state \citep[see, e.g. ][]{Bharali2019, Buisson2019,Zdziarski2021}, soft state \citep[see, e.g. ][]{Fabian2020} and in its final phase \citep[see, e.g. ][]{Xu2020}. 
The truncation of the disk during the hard state is one of the most controversial aspects of the system. On one hand, a number of spectral-timing works pointed out that the disk reaches the ISCO already in hard state, while a contracting lamppost corona is responsible for the hard X-rays emission \citep{Kara2019,Buisson2019,You2021,Wang2021}. On the other hand, a truncated disk scenario was also proposed on the basis of further spectral-timing analyses, even on the same sets of data \citep{Zdziarski2021,DeMarco2021,Axelsson2021}.

\section{Observations \& Data reduction}
In this paper we focus on the spectral analysis of the source in the high luminosity hard state, i.e. from $\approx$ MJD 58190 until $\approx$ MJD 58300. With the aim of modelling broadband X-ray spectra, we included data collected by \swift /XRT, \nustar, \nicer and the Burst Alert Telescope (BAT) onboard \swift. In order to obtain broad-band spectra from observations close in time, we selected only XRT, {\it NuSTAR} and {\it NICER} observations which were quasi-simultaneous, i.e. taken not more than one day apart from each other. Furthermore, in order to avoid any discrepancy between BAT and {\it NuSTAR} due to the spectral variability of the source, we extracted BAT spectra exactly over the duration of each {\it NuSTAR} observation. These criteria led us to narrow down the list of the available {\it NuSTAR} data to eight Epochs. We note that this set of \nustar\ observations has been already analyzed and described by \cite{Buisson2019}. In the following, we will refer to these observations as Epochs and we will label them with numbers from 1 to 8 in chronological order. It is worth noticing that \nicer observations cover five out of eight epochs and that the BAT spectrum in Epoch 6 has been extracted in a longer time interval, due to the low number of visits performed by \swift during the \nustar observation. Details on the selected Epochs are reported in Table \ref{tab:obs}.

\begin{table}
    \centering
    \begin{tabular}{l l l l l }
         \hline
         \hline
         \multicolumn{5}{c}{ XRT}\\
         \hline
         Epoch & ObsID & \multicolumn{2}{c}{Start Time} & Exposure  \\
         & & (UTC) & (MJD) & ks  \\
         \hline
         1 & 00010627001 & 2018-03-14 & 58191.9 & 1.0  \\
         2 & 00010627008 & 2018-03-19 & 58196.8 & 0.98  \\
         3 & 00010627013 & 2018-03-24 & 58201.1 & 0.98  \\
         4 & 00088657001 & 2018-04-04 & 58212.2 & 1.0 \\
         5 & 00010627038 & 2018-04-15 & 58223.3 & 1.7  \\
         6 & 00010627055 & 2018-05-04 & 58240.6 & 1.0  \\
         7 & 00088657004 & 2018-05-18 & 58255.9 & 1.9 \\
         8 & 00088657006 & 2018-06-28 & 58297.3 & 1.8 \\
         \hline
         \multicolumn{5}{c}{{\it NuSTAR} \& BAT$^\dagger$}\\
         \hline
         1 & 90401309002 & 2018-03-14 & 58191.9 & 11.8  \\
         2 & 90401309004 & 2018-03-21 & 58198.0 & 2.8  \\
         3 & 90401309008 & 2018-03-24 & 58201.5 & 3.0  \\
         4 & 90401309012 & 2018-04-04 & 58212.2 & 12.3  \\
         5 & 90401309013 & 2018-04-16 & 58225.0 & 1.8  \\
         6 & 90401309016 & 2018-05-03 & 58241.8 & 13.7  \\
         7 & 90401309019 & 2018-05-17 & 58255.6 & 9.4  \\
         8 & 90401309021 & 2018-06-28 & 58297.2 & 21.4  \\
         \hline
         \multicolumn{5}{c}{ \it   NICER}\\
         \hline
        1 & 1200120103 & 2018-03-13 & 58191.0 & 10.7  \\
        3 & 1200120109 & 2018-03-24 & 58201.0 & 13.0  \\
        4 & 1200120120 & 2018-04-04 & 58212.0 & 6.5  \\
        5 & 1200120130 & 2018-04-16 & 58224.1 & 10.6  \\
        6 & 1200120143 & 2018-05-03 & 58241.2 & 4.0  \\
        \hline
        \hline
    \end{tabular}
    \caption{List of the XRT, {\it NuSTAR} and {\it NICER} observations of the source used in this work. $^\dagger$: BAT survey data were extracted with the same exposure of each {\it NuSTAR} observation, with the exception of Epoch 6, for which a longer ($\sim$ 1 day) time interval was used to increase the statistics.}
    \label{tab:obs}
\end{table}

\subsection{XRT}\label{ss:xrt}
The 2018 outburst of the source was monitored regularly by the XRT telescope on board \swift\ from MJD 58191 (2018, March 14) until MJD 58428 (2018, November 6), with a total of 75 observations performed in Window Timing (WT) mode. The XRT data were first reprocessed with the task \textsc{xrtpipeline}, included in the software package \textsc{HEASOFT} (v. 6.26.1). The source extraction procedure from the cleaned event files was performed with \textsc{ds9}. Since all of the observations had a high count-rate, i.e., always well above the limit for the pile-up correction in WT (100 ct/s), we used an annulus region centered on the source coordinates to extract spectra not affected by pile-up. The outer radius was always chosen as $\sim$47'', while the inner radius of the annulus was selected based on the registered count-rate in accordance with the {\it Swift}/XRT guidelines\footnote{ \url{https://heasarc.gsfc.nasa.gov/lheasoft/ftools/headas/xrtgrblc.html}}. In particular, we used a $\sim$18'' inner radius for Epoch 1, a $\sim$24'' inner radius for Epochs 2, 3, 5, 6 and  a $\sim$28'' inner radius for Epoch 4, 7 and 8. Each spectrum was re-binned with \textsc{grppha} in order to have 150 counts per bin, allowing the use of the $\chi^2$ statistics. \\

\subsection{NuSTAR} \label{ss:nustar}
Data were reduced using the standard \textsc{Nustardas} task, incorporated in \textsc{Heasoft} (v. 6.26.1). We extracted high scientific products (light curves and spectra) using a circular area of 100" radius centered at R.A.\,=\ 18:20:22.0702, and Dec.\,=\ +7:10:58.331, as source region. In order to take into account any background non-uniformity on the detector, we extracted the background spectra using four circles of $\sim$50" radii in different positions with negligible contamination from the source. We then used \textsc{Nuproducts} to build spectra and light curves. We used data from both the two hard X-ray imaging telescopes on board {\it NuSTAR}, i.e. the focal plane mirror (FPM) A and B. The extracted spectra were grouped using the optimal binning recipe by \cite{Kaastra2016} in order to have a grouping which reflects the spectral resolution of the instrument in a given energy range and avoid any oversampling issue. We did not sum the two spectra, but rather fitted them together by leaving a floating cross-normalization constant as suggested by the \nustar team for bright sources\footnote{On the FAQ page, issue 19: \url{https://heasarc.gsfc.nasa.gov/docs/nustar/nustar_faq.html}}. \\

\subsection{BAT}

Data from the BAT survey were also retrieved from the \textsc{HEASARC} public archive. 
The downloaded data were processed using \texttt{BAT-IMAGER} software \citep{Segreto2010}. This code, dedicated to the processing of coded mask instrument data, computes all-sky maps and, for each detected source, produces light curves and spectra. 
Light curves in Crab units were extracted in three energy bands, i.e. 15-40 keV, 40-80 keV, 80-150 keV, with 1 day binning time (Fig. \ref{fig:swift_maxi}).
Spectra were extracted in the range 15-195 keV, with logarithmic binning (for a total of 49 bins) and the official BAT spectral redistribution matrix was used.

\subsection{NICER}
\nicer covered the 2018 outburst of \mysou with almost daily cadence, resulting in hundreds of individual observations (ObsIDs 1200120101-1200120312). In this work, we analysed a sample of the \nicer observations which satisfied the requirement of being quasi-simultaneous, i.e. taken within a day, with respect to the NuSTAR data. The data were reduced using  \textsc{NICER-DAS 2019-05-21} v006. We selected Good Time Intervals using \textsc{NIMAKETIME} and then applied them to the data via \textsc{NIEXTRACT-EVENTS}, selecting events with PI channel between 25 and 1200 (0.25–12.0 keV). In order to fix the distortions due to the \nicer calibration uncertainties, we re-normalised the spectra by using the residuals of a power-law fit to the Crab Nebula \citep{Ludlam2018}. We used the public files \texttt{nixtiaveonaxis20170601v002.arf} and \texttt{nixtiref20170601v001.rmf} as Ancillary Response File and Redistribution Matrix File respectively, retrievable from the {\it NICER} website\footnote{See \url{https://heasarc.gsfc.nasa.gov/docs/nicer/proposals/nicer_tools.html}.}. As a background spectrum we used the public background file \texttt{nixtiback20190807.pi}, also available in the HEASARC archive. 

\section{Spectral analysis}\label{sec:analysis}
\begin{figure}
\centering
\includegraphics[width=\columnwidth]{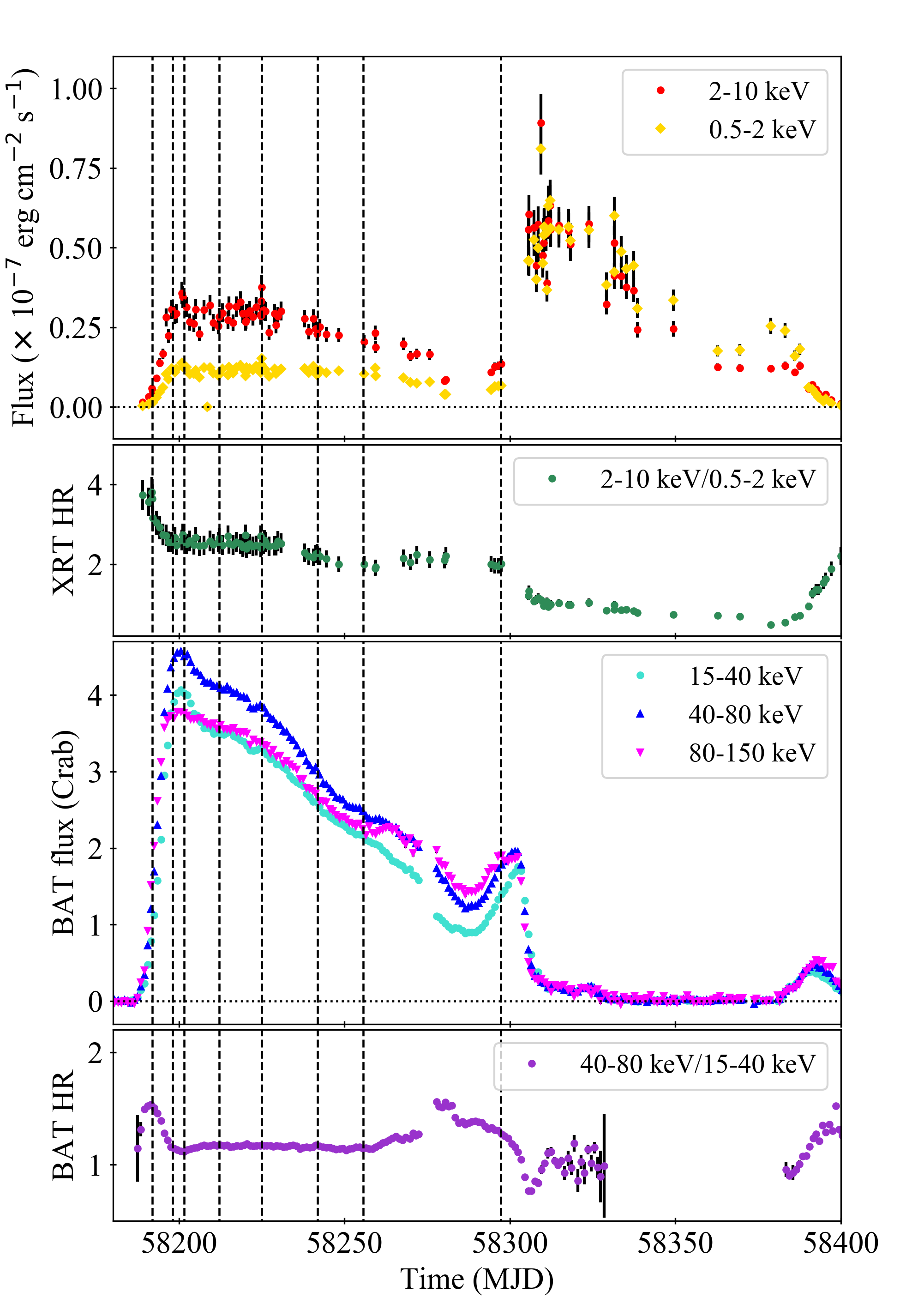}
\caption{{From top to bottom, the \it Swift}/XRT light curves and the related soft X-ray HR, {\it Swift}/BAT light curves and the hard X-ray HR are shown. Energy ranges are reported in the see legends. The {\it NuSTAR} observations are highlighted with black dashed lines.}
\label{fig:swift_maxi}
\end{figure}

\begin{figure}
\centering
\includegraphics[width=1.0\columnwidth]{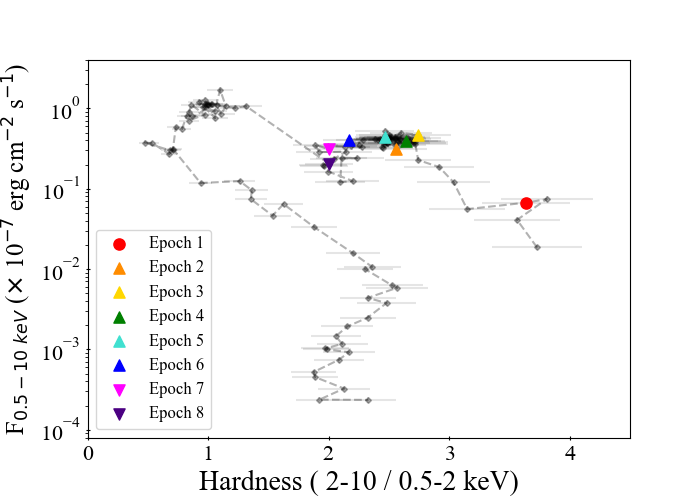}
\caption{XRT HID of the source during the entire outburst. The epochs used in this work are highlighted by the colors and different symbols (Sect. \ref{ss:2refl}), i.e. circle for phase 1 (rise), upper triangles for phase 2 (plateau) and lower triangles for phase 3 (decline). }
\label{fig:hid}
\end{figure}

XRT and BAT light curves and the related hardness ratios are shown in Fig. \ref{fig:swift_maxi}, while in Fig. \ref{fig:hid} the XRT Hardness Intensity Diagram (HID) of the whole 2018 outburst is shown. We used the (absorbed) 0.5-10 keV flux as an indicator of the intensity of the source. In order to estimate the flux, we fitted each XRT spectrum separately in the energy bands 0.5-2 keV and 2-10 keV with a simple power-law (\textsc{POWER} in \textsc{Xspec}) model. We notice that a power-law model, although not being an appropriate choice to describe adequately the spectrum in the whole XRT energy range, approximate satisfactorily its shape in the two energy bands considered. We associated an error equal to 10\% of the estimated value to each of these fluxes. \\ According to their hardness ratio and their timing properties \citep[e.g., ][]{DeMarco2021}, during Epoch 1 the system was in the canonical hard state, while it remained in hard/intermediate state from Epochs 2 to Epoch 8. The latter occurred during a short episode of re-hardening, as better highlighted in the BAT light curve (see Fig. \ref{fig:swift_maxi}). In the following we report on the spectral analysis of these Epochs.
 
The \nustar data usually range from 3-79 keV. However, a remarkable (and unexpected) mismatch between FPMA and FPMB data is sometimes observed below 4 keV, likely due to a known instrumental issue \citep{Madsen2020}. This mismatch was visible in our data. Therefore, we decided not to include data below 4 keV, which are already covered by XRT. Similarly, also \nicer data were ignored below 4 keV, as they showed a soft excess, which could be related to a further Comptonization or a quasi-thermal component, \citep[see fig. 6, ][]{Zdziarski2021}, as well as several lines, both sharp instrumental features \citep{Wang2020} and broad physical lines. A proper investigation of these features goes beyond the purposes of this manuscript and will be reported elsewhere. Moreover, the BAT data showed an unexpected mismatch with \nustar below 30 keV, possibly due to high systematics in this energy range for extremely bright sources. We therefore ignored BAT data below 
30 keV. So that, the final broadband spectra covered the range 0.8-10 keV for XRT, 4-78 keV for \nustar , 4-10 keV for \nicer and 30-190 keV for BAT. \\

Each spectrum was analyzed using the \textsc{tbabs} model in order to take into account the effect of interstellar absorption, with the photoelectric cross sections from \cite{Verner1996} and the element abundances from \cite{Wilms2000}. A \textsc{constant} component was also used to serve as cross-calibration constant. To account for differences in the calibration of {\it NuSTAR} and {\it NICER} \citep[see, also ][]{Zdziarski2021} , we included a cross-calibration function which reads $\textsc{const}\times{E^{\Delta \Gamma}}$, with $\Delta \Gamma$ the discrepancy in photon index between the data sets \citep{Ingram2017,Ursini2020}. $\Delta \Gamma$ was left free to vary for {\it NICER}, while it was fixed to 0 for all the other instruments included in the broadband spectrum. Finally, as each data-set was showing apparent systematics, we applied a 1\% systematic error to the analyzed spectra.

\subsection{Basic ingredients of the JED-SAD spectral model}\label{ss:ingredients}
    The eight X-ray observation were analyzed with the \textsc{JED-SAD} model by applying it to the data in the form of a \textsc{Xspec} table. The key parameters in the model are \citep[but see ][, for a full description ]{Marcel2018b}:
\begin{itemize}
    \item The transition radius, $R_{\rm J}$, between the JED and the SAD, in units of gravitational radii R$_{\rm G}=GM/c^2$.
    \item The mass accretion rate at the transition radius $R_{\rm J}$ or inner mass accretion rate $\dot{M}_{\rm in}$, here expressed in units of Eddington mass accretion rates $\dot{M}_{\rm Edd}=L_{\rm Edd}/c^2$.
    \item The sonic Mach number $m_{\rm S}$ of the accretion flow, defined as the ratio of the mass weighted accretion speed to the sound speed.
    \item The ejection efficiency $p$\footnote{The ejection parameter is labeled $\xi$ in \cite{Marcel2019}. Here we use $p$ to avoid confusion with the ionization parameter of the reflection model.}, which takes into account how effectively the jet extracts angular momentum from the underlying JED, modifying then its structure. Therefore in the JED, the accretion rate $\dot{M} (R)$ \sout{at any radius $R$} scales with $R$ according to the formula $\dot{M} (R) =\dot{M}_{\rm in}\left(\frac{R}{R_{\rm ISCO}} \right)^p$, with $p$ the ejection efficiency. 
    \item The magnetization $\mu$ (see Sect. \ref{ss:jed-sad}).
    \item The fraction $b$ of the accretion power P$_{\rm acc}$ which is released in the JED and ultimately powers the jets, i.e. $b=2 P_{\rm jet}/P_{\rm acc}$.
    \item A geometrical dilution factor $\omega$, corresponding to the fraction of the cold photons emitted by the SAD which cool down (through Inverse Compton) the JED \citep[we refer to ][ and references therein for further discussion on this topic]{Marcel2018b}. In the current version of the model, $\omega$ assumes values in the range 0-0.5.
    \item The radius of the ISCO, $R_{\rm ISCO}$, which coincides with the innermost radius of the JED\footnote{We note however that the JED-SAD calculations have been performed in a Newtonian potential, meaning that the value obtained/chosen for $R_{ISCO}$ needs to be taken with a grain of salt}.
    \item A normalization factor $K_{\rm JEDSAD}$, which determines how the luminosity of the system scales with the distance $D_{\rm kpc}$ (in units of 1 kpc), following the formula: 
\begin{equation}
    K_{\rm JEDSAD}=\left(\frac{10}{D_{\rm kpc}}\right)^2
\end{equation}
\end{itemize}

As explored in detail by \cite{Marcel2018a}, the JED emission in the archetypal object GX 339--4 is best described by setting $\mu$ between 0.1 and 1.0, $b$=0.3, $\omega=0.2$ and $p$=0.01\footnote{This set of parameters, with the exception of $\omega$ that was indeed found through spectral constraints, is also consistent with a full MHD accretion-ejection solution with aspect ratio $\epsilon=H/R\sim0.1$, as expected for a hot JED \citep{Ferreira1997,Petrucci2008}}. In the following we fix $\mu$ to 0.5 and adopt the same set of values for $b$, $\omega$ and $p$ in the case of MAXI J1820+070. Contrarily to the works on GX 339--4, we let $m_{\rm S}$ free to vary in order to investigate its correlation with the high energy cut-off, which can be better constrained in our work due to the \nustar and BAT coverage at hard X-rays. The value of $R_{\rm ISCO}$ in \mysou is not precisely known.  While it has been reported that the ISCO in MAXI J1820+070 could be even lower than 2 R$_{\rm G}$ \citep[e.g. ][]{Kara2019}, it was recently found that the BH in the system is likely slowly spinning \citep{Fabian2020, Zhao2020, Guan2020}, with $a_*$ not higher than 0.5. We chose therefore to use an initial value of $R_{\rm ISCO}$=4 R$_{\rm G}$ (corresponding to a$_*$=0.55), approximately consistent with the estimates reported in these works. The distance of the source was fixed at 3 kpc according to \cite{Atri2020}. We will allow more freedom in $R_{\rm ISCO}$ and $K_{\rm JEDSAD}$ and discuss the dependence of our results on the choice of these parameters in Sect. \ref{ss:rin}-\ref{ss:distance}. A summary of the parameters in the model and the values adopted in this work is presented in Table \ref{tab:jedsad}. \\

\begin{table} \label{tab:jedsad}
\centering
\begin{tabular}{l  l  l l}
\hline 
\hline
Parameter & &  Description & Adopted  \\
 & & & Value(s) \\
\hline
$R_{\rm J}$ & (R$_{\rm G}$) & Transition radius & (free) \\
$\dot{M}_{\rm in}$ & $\left(\dot{M}_{\rm Edd}\right)$ & Inner mass-accretion & (free) \\
& & rate & \\
$p$ & & Ejection efficiency & 0.01$^a$ \\
$\mu$ & & Magnetization & 0.5$^a$ \\
$b$ & & Jet power & 0.3$^a$ \\
$\omega$ & & Dilution factor & 0.2$^a$ \\
$m_{\rm S}$ & & Sonic Mach number & (free) \\
$R_{\rm ISCO}$ & (R$_{\rm G}$) & ISCO & 4, 2 \\
$D_{\rm kpc}$ & (kpc) & Distance in kpc & 3$^b$ \\ 
\hline
\hline
\end{tabular}
\caption{Parameters used in the \textsc{JED-SAD} model. $^a$: the parameter was set to the same value found by \cite{Marcel2018a}. $^b$: in accordance with \cite{Atri2020}. }
\label{tab:jedsad}
\end{table}

\subsection{The two reflection models}\label{ss:2refl}
\begin{figure}
\centering
\includegraphics[width=0.9\columnwidth]{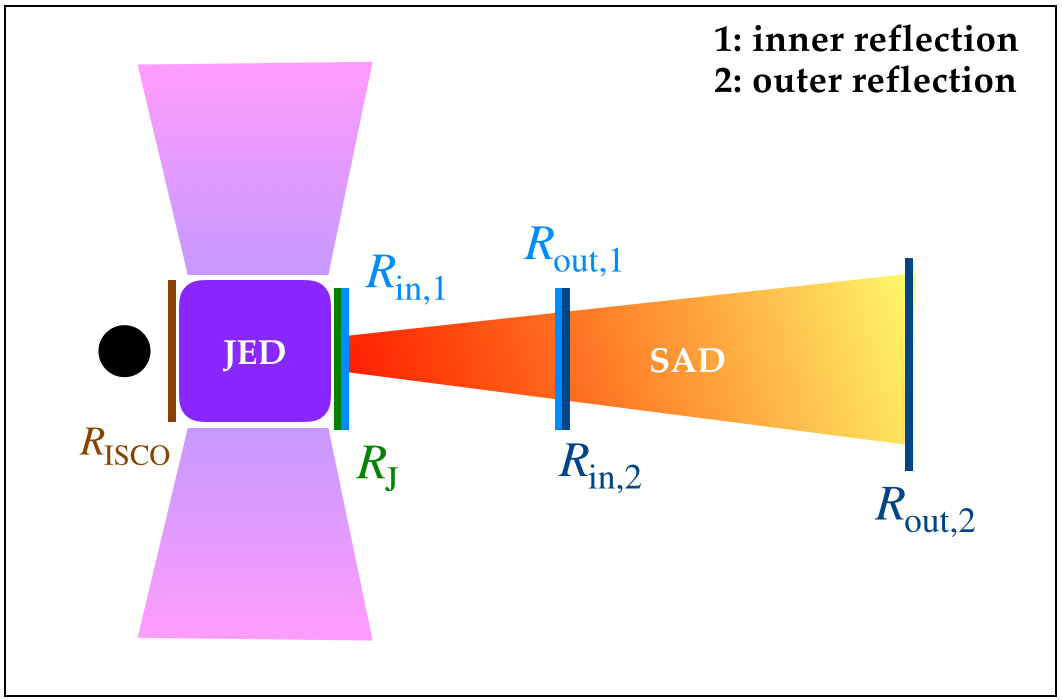}
\caption{A sketch representing the proposed two-reflection geometry underlying Model 2. From left to right: the black hole (in black), the inner JED (violet) and the jets (lilac), the outer SAD (from red to yellow).}
\label{fig:sketch}
\end{figure}

A first analysis without any reflection component results in extremely poor fits and strong residuals in the iron line and Compton hump regions. This was expected because reflection is not taken into account in the JED-SAD model. We therefore used a reflection table component, that we will label \textsc{refl} for simplicity, based on the \textsc{xillver} reflection model \citep{Garcia2013}. This table has been produced by simulating spectra for different combinations of $R_{\rm J}-\dot{M}_{\rm in}$ pairs, fitting them with a simple cut-off power-law model and finally injecting the corresponding spectral index and high energy cut-off in the \textsc{xillver} table. More details on this model will be given in Barnier et al. ({\it submitted}). 
The parameters of this component are: the iron abundance $A_{\rm Fe}$, the disk ionization parameter $\log \xi$ and a normalization $K$.  
Since \textsc{refl}  does not take into account the relativistic blurring effects, we applied the convolution kernel \textsc{kdblur}, which smears the reflection spectrum according to the original calculations by \cite{Laor1991}. This model considers four additional parameters: the emissivity $\epsilon$, the inner and outer radius of the reflecting disk $R_{\rm in}$ and $R_{\rm out}$, the inclination $i$. In order to reduce the number of degrees of freedom we set $\epsilon$ to the reasonable value of 3 \citep[see, e.g. ][]{Dauser2013,Xu2020} and $i$ to 70$^\circ$ \citep[consistent with the estimated source inclination, see e.g. ][]{Kajava2019}. As the reflection component is produced by the disk/SAD surface reprocessing the hard X-ray photons emitted by the corona/JED, within the \textsc{JED-SAD} framework the inner radius of the reflection $R_{\rm in}$ should coincide with the transition radius, $R_{\rm J}$, between JED and SAD. In all spectral fits, $R_{\rm J}$ and $R_{\rm in}$ were therefore linked together. \\
The first model we applied, Model 1, reads:
\begin{multline}
    {\rm \bf Model \ 1: } \textsc{tbabs}\times(\textsc{atable}(\textsc{jedsad}) + \\ +\textsc{kdblur}\times\textsc{atable}(\textsc{refl}))
\end{multline}
It is noteworthy that two reflection components were already suggested by \cite{Kara2019} on the basis of reverberation time lags study with \nicer data. Thereafter a number of authors used a double-reflection model in the spectral analysis of the same {\it NuSTAR} observations presented here \citep{Buisson2019,Chakraborty2020,Zdziarski2021}. They all found both a low ionisation reflection component and a highly ionised inner reflection component, which contribute to the observed iron line profile providing respectively a narrow core and a broad base. Therefore, we also tested a second model, Model 2, where a second \textsc{refl} component has been added:
\begin{multline}
     \rm{\bf Model \ 2: } \textsc{tbabs}\times(\textsc{atable}(\textsc{jedsad}) + \\ + \textsc{kdblur}\times\textsc{atable}(\textsc{refl})+\textsc{kdblur}\times\textsc{atable}(\textsc{refl}))
\end{multline}
When Model 2 is applied, we use the subscripts '1' and '2' to refer to the parameters of the inner and outer reflection components, respectively. The purpose of this model is to reproduce a physical scenario where reflection comes from two distinct regions, characterized by different ionisation parameters and therefore presumably placed at different distances from the illuminating corona (the JED here). Therefore we tied $R_{\rm in,1}$ to $R_{\rm J}$ and $R_{\rm out,1}$ to $R_{\rm in,2}$, in order to place the two reflection regions at separate but neighboring regions\footnote{This is only a working assumption, as it cannot be excluded that regions may not be exactly adjacent to each other.}, as shown in Figure \ref{fig:sketch}. The outer radius of the remote reflection is instead left free to vary. The normalisation and ionization parameters were left free as well in both reflection components, so that we have two different normalization and ionization values: $K_1$, $K_2$, $\xi_1$ and $\xi_2$. All the other parameters, i.e. the inclination, the iron abundance and the emissivity, were tied between the two reflection components.

\subsection{Spectral fits with $R_{\rm ISCO}$=4 R$_{\rm G}$}\label{ss:analysis}
\begin{figure*}
\centering
\includegraphics[width=\textwidth]{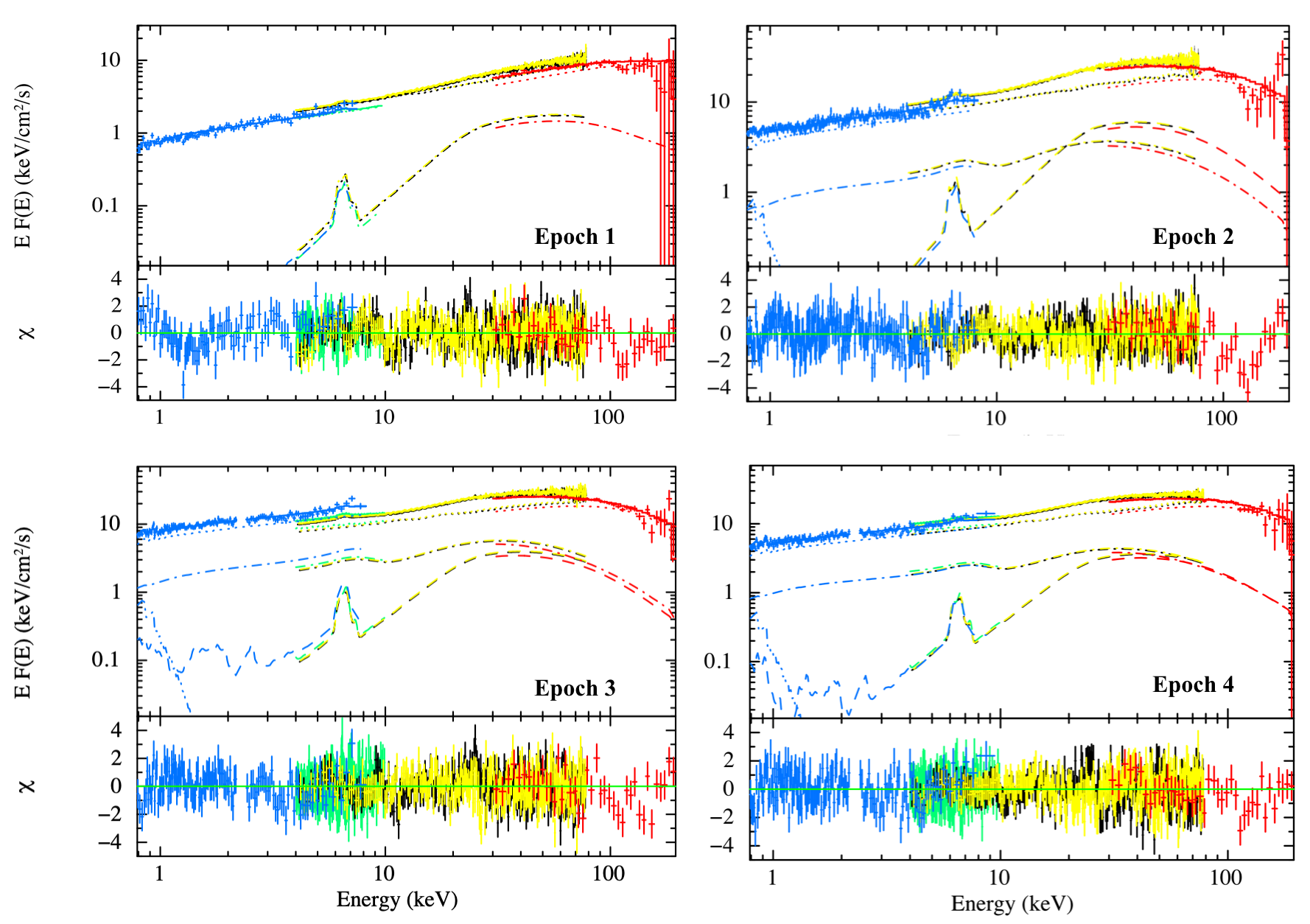}
\caption{Energy spectra with the best-fit models, i.e. Model 1 for Epoch 1 and Model 2 for Epochs 2-4, and residuals. Data: XRT (blue), {\it NICER} (green), {\it NuSTAR} (yellow-black) and BAT (red). Different linestyles were adopted to distinguish between the different components, in particular: dot for \textsc{JED}, dash-dot-dot-dot for \textsc{SAD}, dash-dot for inner \textsc{refl}, dash for the outer \textsc{refl}.}
\label{fig:residuals1}
\end{figure*}

\begin{figure*}
\centering
\includegraphics[width=\textwidth]{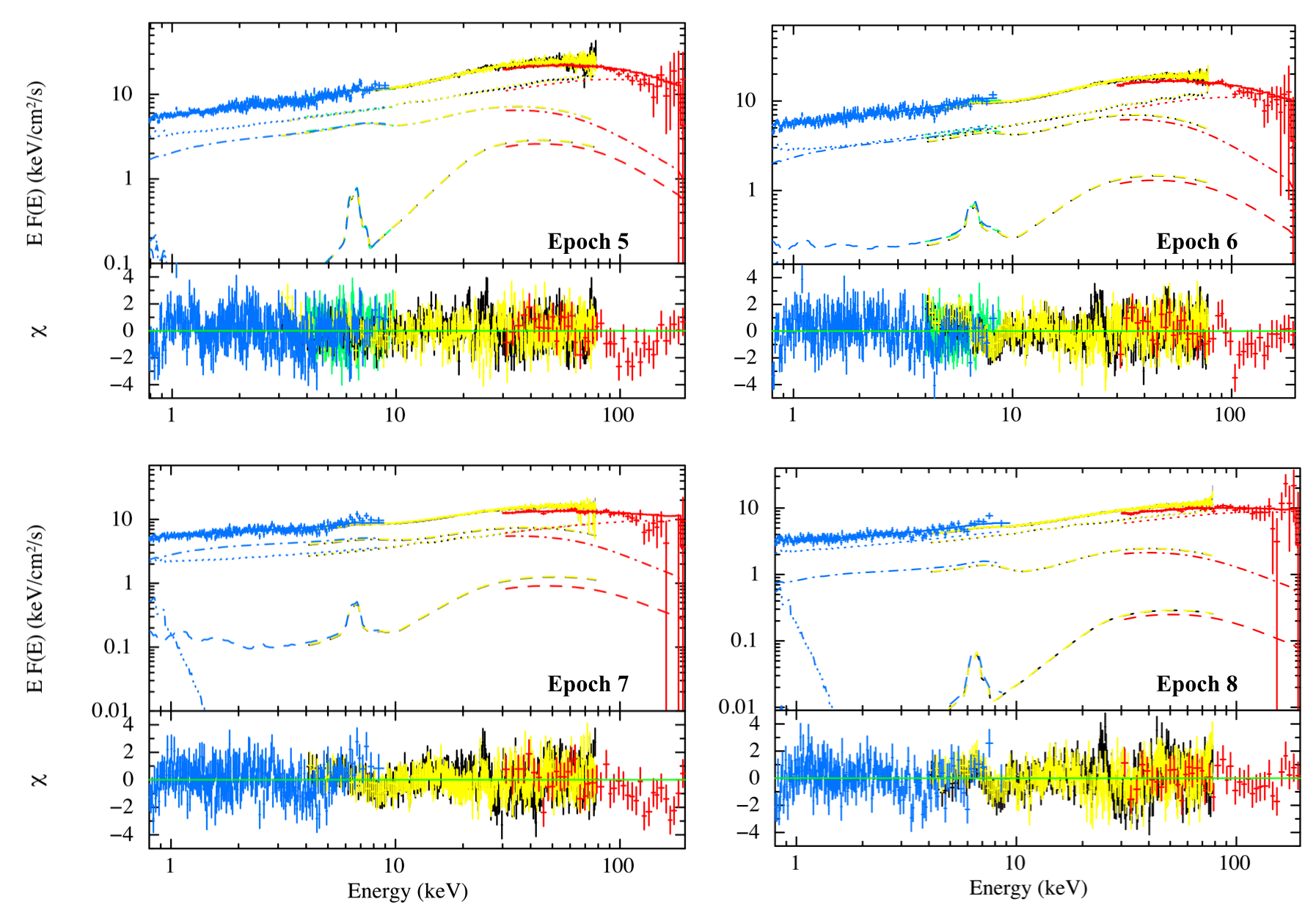}
\caption{Energy spectra, best-fit model (Model 2) and residuals for Epochs 5-8. Data: XRT (blue), {\it NICER} (green), {\it NuSTAR} (yellow-black) and BAT (red). Different linestyles were adopted to distinguish between the different components, in particular: dot for \textsc{JED}, dash-dot-dot-dot for \textsc{SAD}, dash-dot for inner \textsc{refl}, dash for the outer \textsc{refl}.}
\label{fig:residuals2}
\end{figure*}

Based on the fit results described below, we can divide our observations in three groups or "phases": phase 1 corresponds to Epoch 1, phase 2 includes observations from Epoch 2 to Epoch 6 , phase 3 which corresponds only to Epochs 7 and 8. The grouping of these Epochs is consistent with the XRT HID (Fig. \ref{fig:hid}), where an analogous clustering of observations in three regions can be observed. A similar subdivision was also proposed by \cite{DeMarco2021}, with phases labeled as "Rise", "Plateau" and "Bright decline", respectively. In the following, we will treat these phases separately. In all phases, we did not manage to constrain the values of $R_{\rm in, 2}$ and $R_{\rm out, 2}$, as the fit was basically insensitive to variations of these parameters. We therefore set $R_{\rm in, 2}$ to 300 R$_{\rm G}$ and $R_{\rm out,2}$ to 10$^4$ R$_{\rm G}$. We found values of $A_{\rm Fe}$ always between 2 and 3 times the solar abundance\footnote{We caution the reader that the incompatibility between the best-fit value of $A_{\rm Fe} = 2.0-2.4$ found for Epoch 2 and the common value of about 3 found for the other seven epochs, does not have a physical origin. For consistency, we tried fixing $A_{\rm Fe}$ to a value of 3 also in Epoch 2, but the fit, and particularly the iron line residuals, significantly worsened. This may be due to systematics in \nustar between 8 and 11 keV which interfere with a correct modelisation of the iron line region.}. \\
The final results are reported on Table \ref{tab:xrtnusttworin4}, while corresponding best-fit models and residuals are shown in Figure \ref{fig:residuals1}-\ref{fig:residuals2}.

\begin{table*}
\centering
\begin{tabular}{l  l  l l l l }
\hline 
\hline
\multicolumn{5}{l}{\textbf{Spectral analysis in the case of $R_{\rm ISCO}$=4 R$_{\rm G}$}} \\
\hline
Epochs & & 1 & 2 & 3 & 4 \\
\cmidrule(l){3-6}
N$_H$ & $\times$10$^{22}$ cm$^{-2}$ & $0.560 \pm 0.020$ & $0.190\pm0.020$  & $0.192^{+0.014}_{-0.015}$ & $0.150\pm0.020$  \\
$R_{\rm J}$ & R$_{\rm G}$ &  $56.8^{+1.0}_{-1.1}$ & $35.9^{+5.0}_{-1.4}$ & $44.8^{+0.3}_{-0.5}$ & $43.9^{+1.1}_{-1.0}$ \\
$\dot{M}_{\rm in}$ & $\dot{M}_{\rm Edd}$ & $0.838^{+0.008}_{-0.009}$  & $1.760\pm0.080$  & $2.360^{+0.070}_{-0.050}$  & $2.120^{+0.060}_{-0.030}$  \\
$m_{\rm S}$ & & $1.301^{+0.017}_{-0.016}$ & $1.250^{+0.020}_{-0.050}$ & $1.243\pm0.001$  & $1.250^{+0.020}_{-0.010}$ \\
$A_{\rm Fe}$ & & $2.9^{+0.3}_{-0.4}$ & $2.2^{+0.2}_{-0.2}$ & \multicolumn{2}{c}{(3)}  \\
$R_{\rm in,1}$ & $R_{\rm G}$ & \multicolumn{4}{c}{=$R_{\rm J}$} \\
$R_{\rm in, 2}$ & $R_{\rm G}$ & \ & \multicolumn{3}{c}{(300)} \\
$R_{\rm out,1}$ & $R_{\rm G}$ & 10$^4$ & \multicolumn{3}{c}{=$R_{\rm in, 2}$} \\
$R_{\rm out, 2}$ & $R_{\rm G}$ & \  & \multicolumn{3}{c}{(10$^4$)} \\
$i$ & $^\circ$ & \multicolumn{4}{c}{(70)}  \\
$\log{\xi_{1}}$ & & $2.22^{+0.13}_{-0.23}$ & (3.50)  & $3.71^{+0.02}_{-0.08}$ & $3.69^{+0.04}_{-0.13}$  \\
$\log{\xi_{2}}$ & & \  & <1.70 & <1.65 & $2.07^{+0.22}_{-0.01}$ \\
K$_1$ & ($\times$10$^{-6}$) & $42.0^{+17.0}_{-20.0}$ & $3.3^{+1.0}_{-0.9}$ & $3.6^{+0.4}_{-0.2}$  & $3.0^{+0.9}_{-0.4}$ \\
K$_2$ & ($\times$10$^{-4}$) & \ & $232.0^{+15.0}_{-22.0}$   & $0.6^{+0.1}_{-0.2}$ & $1.3\pm0.5$ \\
\hline
$\chi^2_{\nu}$ & (d.o.f.) & 1.01(746) & 1.03(755) & 1.00(1036) & 0.99(1029) \\
\hline
\\
\hline
Epochs & & 5 & 6 & 7 & 8 \\
\cmidrule(l){3-6}
\textbf{N$_H$} & $\times$10$^{22}$ cm$^{-2}$ & $0.13\pm0.01$  & $0.11\pm0.01$ & $0.15\pm0.02$  & $0.15\pm0.02$ \\
$R_{\rm J}$ & R$_{\rm G}$ &   $47.8^{+3.4}_{-2.1}$ & $45.8^{+1.3}_{-1.1}$ & $33.6^{+0.6}_{-0.5}$ & $33.3^{+1.2}_{-0.7}$ \\
$\dot{M}_{\rm in}$ & $\dot{M}_{\rm Edd}$ & $1.56^{+0.07}_{-0.10}$ & $1.25^{+0.02}_{-0.04}$  & $1.00^{+0.05}_{-0.03}$  & $1.06^{+0.02}_{-0.04}$ \\
m$_{\rm  S}$ & & $1.25^{+0.02}_{-0.01}$ & (1.25) & >1.49 & >1.49 \\
$A_{\rm Fe}$ & & \multicolumn{3}{c}{(3.0)} & $3.4^{+0.3}_{-0.4}$  \\
$R_{\rm in,1}$ & $R_{\rm G}$ & \multicolumn{4}{c}{=$R_{\rm J}$} \\
$R_{\rm in, 2}$ & $R_{\rm G}$ & \multicolumn{4}{c}{(300)} \\
$R_{\rm out,1}$ & $R_{\rm G}$ & \multicolumn{4}{c}{=$R_{\rm in, 2}$} \\
$R_{\rm out, 2}$ & $R_{\rm G}$ & \multicolumn{4}{c}{(10$^4$)} \\
$i$ & $^\circ$ & \multicolumn{4}{c}{(70)}  \\
$\log{\xi_{1}}$ & & $3.81^{+0.06}_{-0.03}$  & $3.80^{+0.05}_{-0.03}$ & $3.79^{+0.03}_{-0.02}$ & $3.48^{+0.13}_{-0.09}$  \\
$\log{\xi_{2}}$ & & $2.32^{+0.04}_{-0.19}$ & $3.04^{+0.07}_{-0.10}$ & $2.99^{+0.05}_{-0.12}$ & (2.00) \\
K$_1$ & ($\times$10$^{-6}$) & $3.5\pm0.3$ & $3.9^{+0.7}_{-0.8}$ & $3.6\pm0.4$ & $2.7\pm 0.6$ \\
K$_2$ & ($\times$10$^{-5}$) & $0.70^{+0.50}_{-0.20}$ & $0.63^{+0.38}_{-0.10}$  & $0.61^{+0.18}_{-0.07}$ & 0.15$^{+0.08}_{-0.12}$  \\
\hline
$\chi^2_{\nu}$ & (d.o.f.) & 0.98(1146) & 1.01(991) & 1.12(794) & 1.15(748) \\
\hline
\hline
\end{tabular}
\caption{Fit results for all epochs with Model 1 for epoch 1 and Model 2 for the remaining epochs. Quoted errors reflect 90\% confidence level. The parameters which were kept frozen during the fits are reported between round parentheses. }
\label{tab:xrtnusttworin4}
\end{table*}

\paragraph{Phase 1: The system in the  hard state with one reflection component}\mbox{}\\
Epoch 1 is satisfactorily fitted with both Model 1 and Model 2, with $\chi^2_\nu$ (d.o.f.) of 1.01(749) and 0.90(748), respectively. In both cases, the SAD is truncated far away from the BH, i.e. with $R_{\rm J}$ of $\sim$60 R$_{\rm G}$ and $\sim$ 150 R$_{\rm G}$ for Model 1 and Model 2, respectively. However, with Model 2, the value obtained for $\log{\xi}_1$ goes beyond the threshold of 4, i.e. around 4.3. We consider such high value as unphysical for several reasons. First of all, such a highly ionized medium at the edge of the disk is hard to reconcile with it being truncated at 150 R$_{\rm G}$. Secondly, we note that with the very high value found for $\log{\xi_1}$, the inner reflection component has almost no significant fluorescence line and basically serves as an additional continuum component. Finally, the spectral shape of the reflection component computed with \textsc{xillver} for $\log{\xi}$>4 is unreliable at high energies\footnote{J. Garcia, private communication.}. According to these arguments, the presence of the second reflection component results most likely spurious for Epoch 1. Using Model 1, $\log{\xi}$\footnote{The subscript is removed here since in Model 1 there is no need to distinguish between component "1" and component "2".} is found in the range 2.0-2.4, which is instead more plausible. 

\paragraph{Phase 2. The outer reflection component arises}\mbox{}\\
Epochs from 2 to 6 are poorly fitted by Model 1, with $\chi^2_\nu$ values all above 1.25 and, most importantly, evident unmodelled structures between 6 and 10 keV in the residuals. Instead, the application of Model 2 results in good fits and quite "flat" residuals. The picture captured by the fits is similar for the epochs in this group. The best-fit value of $R_{\rm J}$ decreases significantly with respect to the previous phase, i.e. attaining values in the range 35-50 R$_{\rm G}$. Simultaneously, $\dot{M}_{\rm in}$ evolves substantially, going from 1.8 $\dot{M}_{\rm Edd}$ (Epoch 2) to 2.4 $\dot{M}_{\rm Edd}$ (Epoch 3), and then decreases again to 1.3 in epoch 6. $m_{\rm S}$ retains a well-determined value of about 1.25 for all the epochs in this phase, with the only exception of epoch 6, where the parameter was basically unconstrained and therefore fixed. In all epochs, ${\xi_1}$ is higher than ${\xi_2}$ by at least an order of magnitude. Interestingly, while $\xi_1$ is quite stable, an increasing trend can be observed in $\log{\xi_2}$, which goes from an upper limit of 1.7 in epoch 2, to 3 in epoch 6. The possibility of this trend being related to an evolution of the inner radius of the outer reflection will be explored in the Discussion (see Section \ref{sec:discmaxi}).

\begin{figure*}
\centering
\includegraphics[width=0.9\columnwidth]{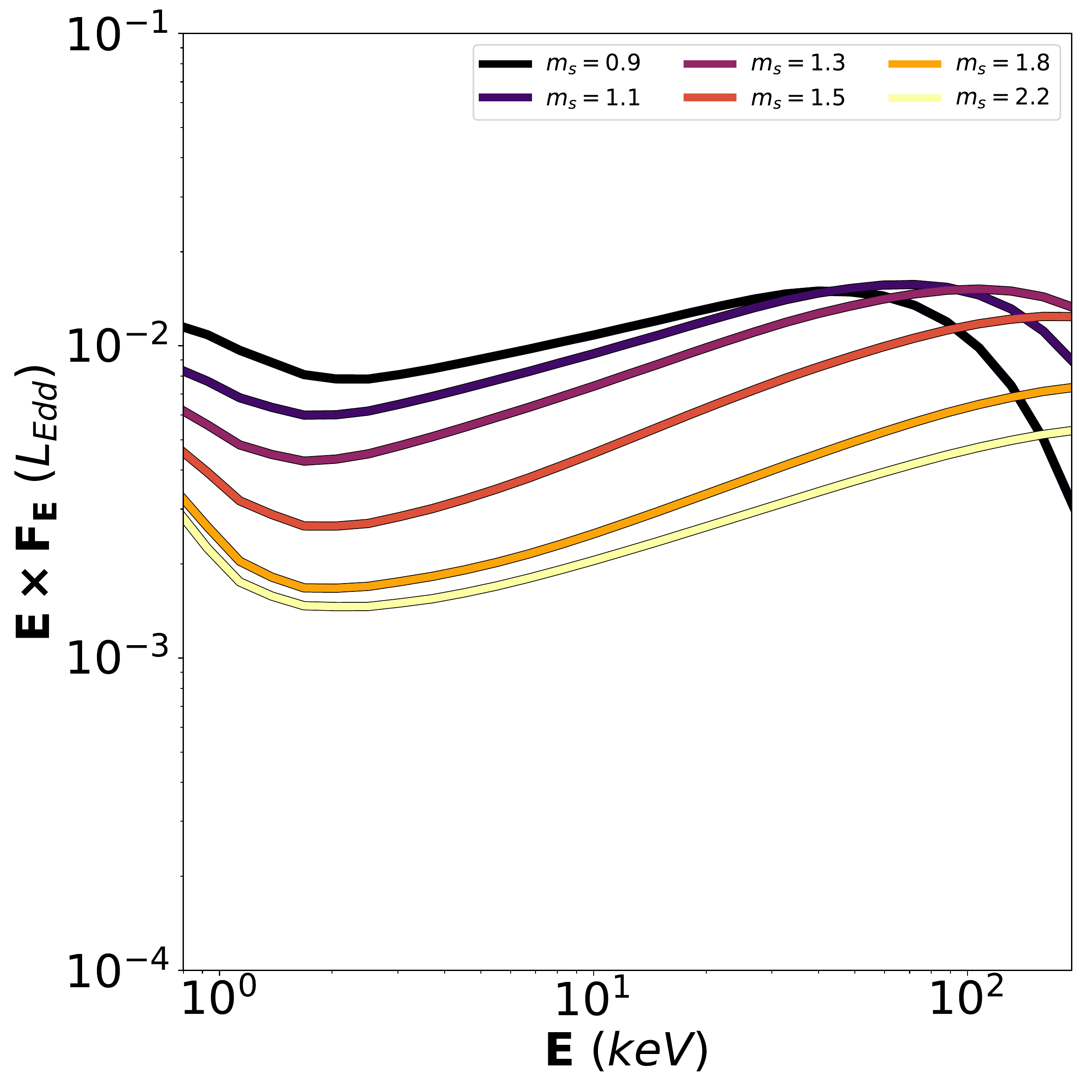}
\includegraphics[width=0.9\columnwidth]{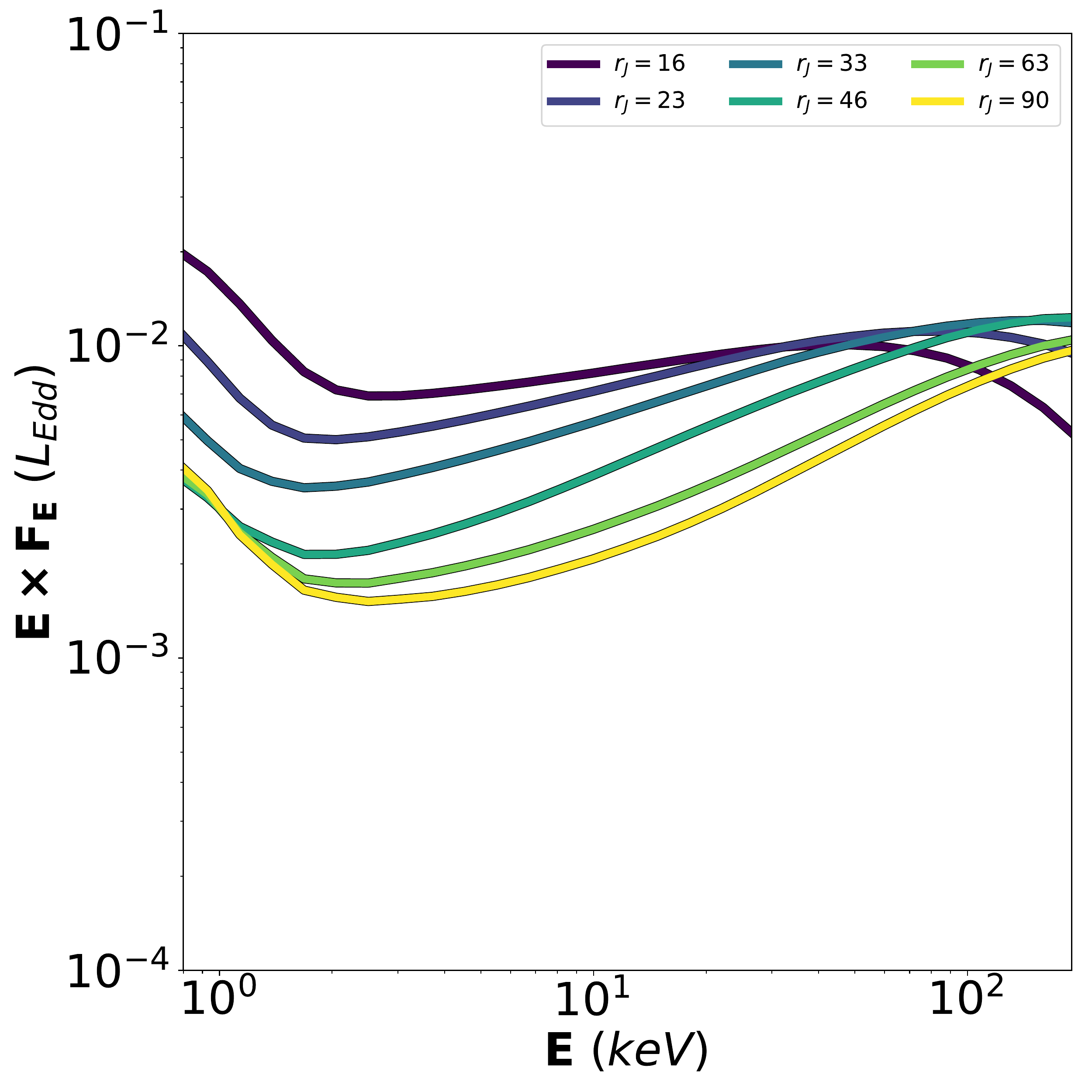}
\caption{JED solutions showing the impact of $m_{\rm S}$ and $R_{\rm J}$ on the high energy cut-off. In the {\it Left} panel, we show different JED solutions obtained keeping $R_{\rm J}$ to 40 R$_{\rm G}$ for different values of $m_{\rm S}$. In the {\it Right} panel, the different JED solutions are produced for different values of $R_{\rm J}$, here expressed as $r_{\rm J}=R_{\rm J}/{\rm R}_{\rm G}$, and $m_{\rm S}$ fixed to 1.5. The plot is used to display how increasing both $m_{\rm S}$ and increasing $R_{\rm J}$ pushes the hard X-rays cut-off to higher energies.}
\label{fig:ms-rj}
\end{figure*}

\begin{figure}
\centering
\includegraphics[width=\columnwidth]{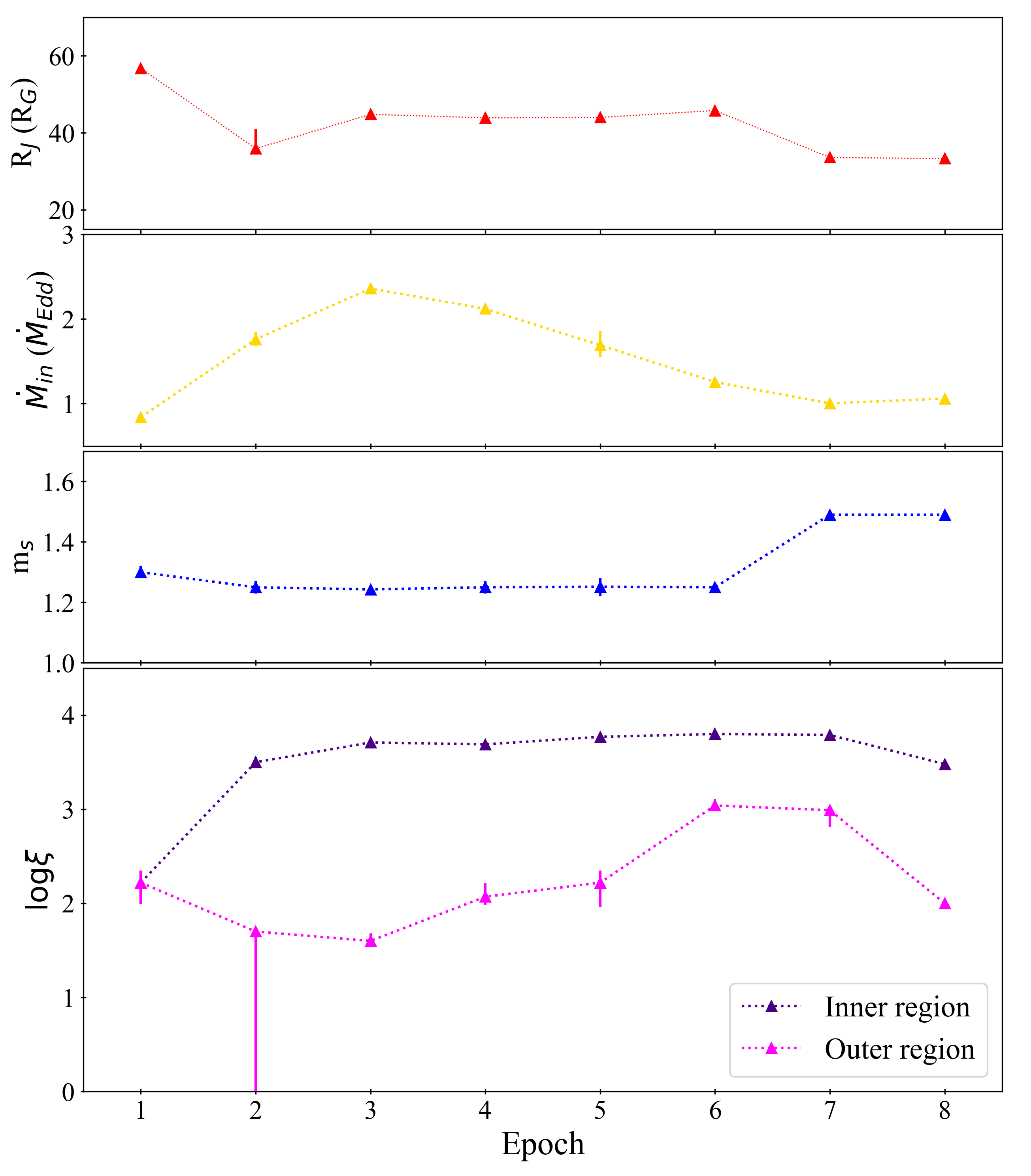}
\caption{Evolution of the best-fit parameters for $R_{\rm J}$, $\dot{M}_{\rm in}$ and $m_{\rm S}$ over the eight analyzed Epochs. In all panels, triangles and bold lines are used for fits performed with $R_{\rm ISCO}$ at 4 R$_{\rm G}$. The reported errors are likely underestimated and an additional, about 10\%-20\%, systematic error has to be considered to account for the uncertainty on $\omega$ and $K_{\rm JEDSAD}$ (see Sect. \ref{ss:distance}).}
\label{fig:threeparam}
\end{figure}

\begin{figure}
\centering
\includegraphics[width=\columnwidth]{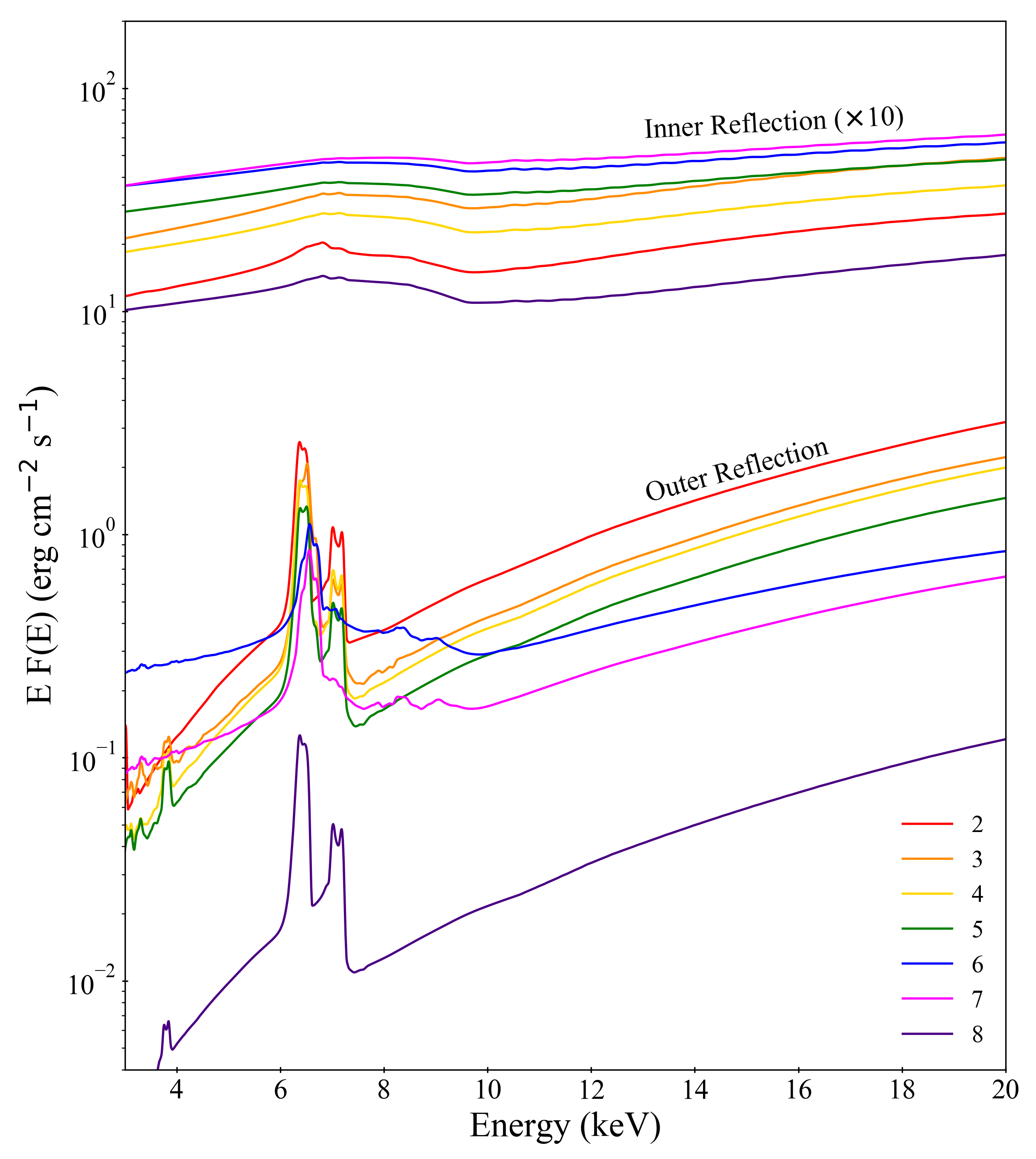}
\caption{Best-fit models for inner and outer reflection in Epochs 2-7. Inner reflection models have been multiplied by a factor 10 for clarity.}
\label{fig:reflection_profiles}
\end{figure}

\paragraph{Phase 3: The role of $m_{\rm S}$ in the re-hardening}\mbox{}\\
Model 1 also gives an unacceptable fit for Epoch 7, but is instead compatible with Epoch 8, with a $\chi^2_\nu$ (d.o.f.) of 1.15 (750). Model 2 gives a good fit to Epoch 7 and an even better fit to Epoch 8, with a probability of improvement by chance of $\sim$ 10$^{-5}$ with respect to Model 1. We then conclude that a double reflection scenario is also statistically preferred in phase 3. As in the previous phase, $R_{\rm J}$ decreases again to about 33-34 R$_{\rm G}$, while $\dot{M}_{\rm in}$ is constrained at $\sim$ 1 $\dot{M}_{\rm Edd}$. The most apparent evolution with respect to the previous phase consists in a re-hardening of the spectrum, as witnessed by the high energy cut-off of the BAT data moving towards higher energies (see the BAT light curve, Fig. \ref{fig:swift_maxi}). Since some level of degeneracy between $R_{\rm J}$ and $m_{\rm S}$ clearly exists (see Fig. \ref{fig:ms-rj}), such trend can be explained by an increase in both these parameters. For example, Epoch 7 can be fitted keeping $m_{\rm S}$ fixed to 1.3, in consistency with the previous phase, and with $R_{\rm J}$ of 38-39 $R_{\rm G}$ or leaving $m_{\rm S}$ free, a case where $m_{\rm S}$ attains a value of 1.5 and $R_{\rm J}$ decreases to 33-34 R$_{\rm G}$. Both fits are acceptable, but keeping $m_{\rm S}$ fixed results in slightly unmodelled residuals in the iron line region, contrarily to the case where $m_{\rm S}$ is free. We therefore conclude that the spectral hardening in this phase is not reflected in the values of $R_{\rm J}$, which actually suggests that the inner edge of the disk keeps moving inwards as expected during the transition to the soft state, but seems to be instead produced by an increase in $m_{\rm S}$, producing a more rarefied and hotter accretion flow in the JED. Another interesting evolution can be found in the values of $\xi_{1}$ and $\xi_2$: while Epoch 7 has values compatible with Epoch 6 (and phase 2 in general), in Epoch 8 both values $\log{\xi_1}$ and $\log{\xi_2}$ decrease significantly, i.e. to values of $\sim$ 3.5 and 2.0 (which was fixed since it was left unconstrained by the fit), respectively. \\
A summary on the evolution of the main parameters at play is presented in Fig. \ref{fig:threeparam}. We also show the evolution of the reflection components in Fig. \ref{fig:reflection_profiles}.

\begin{figure*}
\centering
\includegraphics[width=\textwidth]{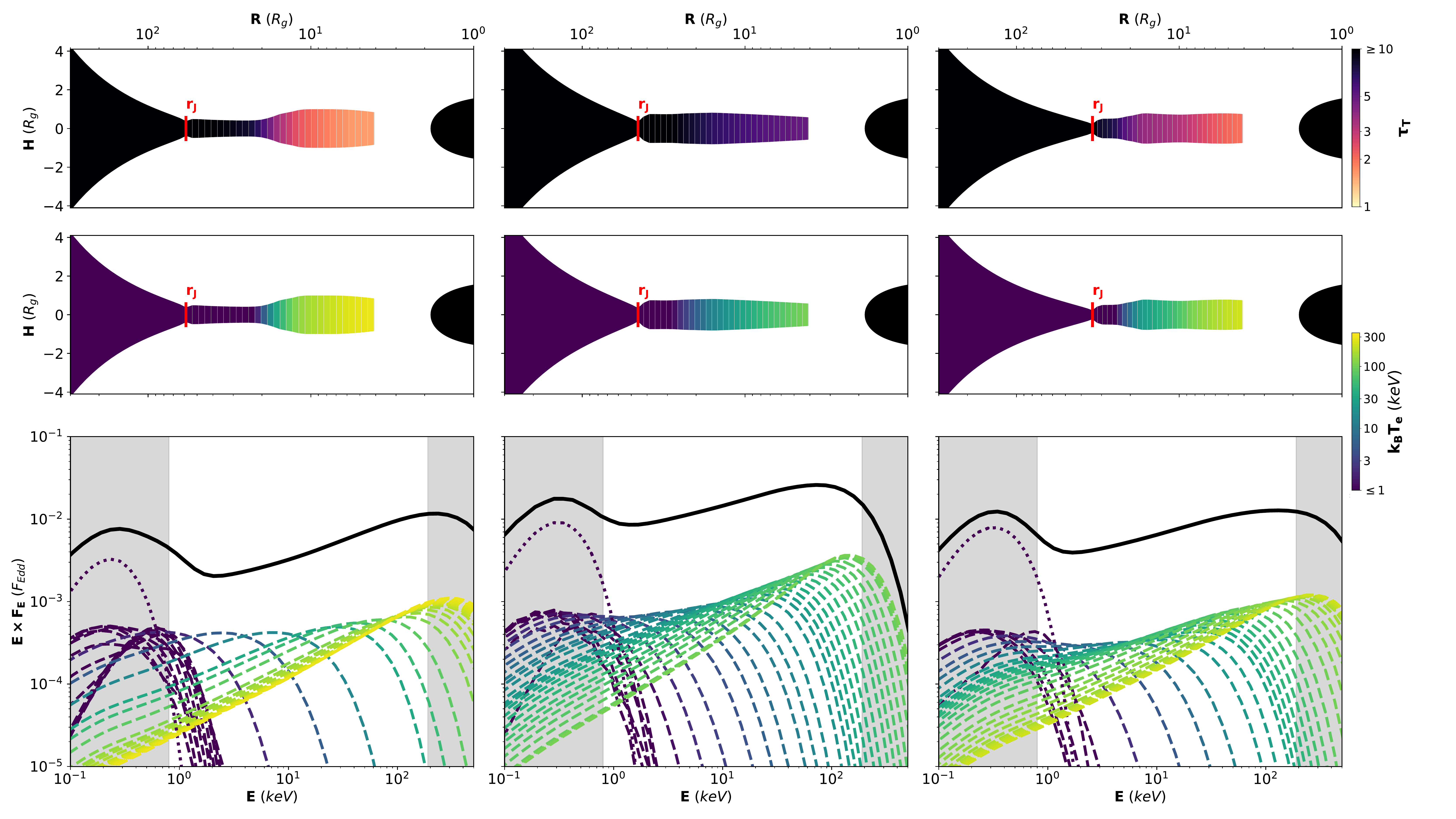}

\caption{JED-SAD solutions for Epochs 1, 5 and 8, used to represent phases 1, 2 and 3, respectively. In order to calculate these diagrams, we used the best-fit parameters for $R_{\rm J}$, $\dot{M}_{\rm in}$ and $m_{\rm S}$ reported in Table \ref{tab:xrtnusttworin4}. In particular, for phase 1 we used $R_{\rm J}$=57 R$_{\rm G}$, $\dot{M}_{\rm in}$=0.8 $\dot{M}_{\rm Edd}$ and $m_{\rm S}$=1.3, for phase 2 we used $R_{\rm J}$=44 R$_{\rm G}$, $\dot{M}_{\rm in}$=1.7 $\dot{M}_{\rm Edd}$ and $m_{\rm S}$=1.3 and for phase 3 the values adopted were $R_{\rm J}$=33 R$_{\rm G}$, $\dot{M}_{\rm in}$=1.1 $\dot{M}_{\rm Edd}$ and $m_{\rm S}$=1.5. In all panels, the radial distribution is divided in 31 portions (annuli), 30 for the JED and one for the SAD. In the top and middle panels, we present how the actual shape $z=H/R$ of the disk evolves with $R$. The distribution of the Thomson optical depth ($\tau_{\rm T}$, top-right color bar) and of the electron temperature ($kT_{\rm e}$, right color bar) is displayed with colors in the top and middle panels, respectively. In the same plots, the event horizon of the BH is represented by a half black circle. The total emitted spectrum is reported in the bottom panels with a solid black line. The contribution from each of the 31 considered annuli is highlighted with a dotted line for the SAD region and dashed lines for the JED regions. The sections of the plot corresponding to the energy ranges that are not covered by XRT, \nicer, \nustar\ and BAT data are shown with grey background.}
\label{fig:phases}
\end{figure*}

Using the best-fit values found for $R_{\rm J}$, $\dot{M}_{\rm in}$ and $m_{\rm S}$, it is possible to produce diagrams for the main physical properties of the corresponding hybrid JED-SAD configuration. These diagrams are presented in Fig. \ref{fig:phases} for Epoch 1, 5 and 8, each of them chosen to represent one of the three phases. In these plots, the accretion flow has been divided into 31 radial zones, 30 for the JED, one for the SAD. 
The scale height $H$, the Thomson optical depth $\tau_{\rm T}$, the electron temperature $kT_{\rm e}$ and the emitted spectrum were calculated and plotted in Fig. \ref{fig:phases} for each one of these 31 regions \citep[see Appendix C in ][for further details on how these plots were produced]{Ursini2020}. The plots show that the increase in $\dot{M}_{\rm in}$ in phase 2 corresponds to a geometrically thinner, optically thicker and colder JED with respect to phase 1. The evolution of the accretion flow is also apparent from the softening of the spectra emitted by the single rings forming the JED. The re-hardening episode in phase 3 is witnessed by the slight increase in the spectral high energy cut-off combined with an increase in both $kT_{\rm e}$ and $\tau_{\rm T}$.

\subsection{Spectral fits with $R_{\rm ISCO}$=2 R$_{\rm G}$}\label{ss:rin}  

In Sect. \ref{ss:analysis}, we kept the innermost radius $R_{\rm ISCO}$ frozen to 4 R$_{\rm G}$, corresponding to a moderately spinning black hole, as apparently suggested by recent spectral analysis \citep{Fabian2020,Zhao2020,Guan2020}. In this section, we replicated the previous fitting procedure after setting $R_{\rm ISCO}$ to 2 R$_{\rm G}$, i.e. corresponding to a rapidly spinning BH (a$_* \sim$ 0.95). 

In this new ensemble of fits, a subdivision in three phases can again be individuated. Furthermore, a double reflection model is still necessary to fit the Epochs, with the exception of Epoch 1, for which Model 2 results again in oddly high values of $\log{\xi}_1$. However, a lower $R_{\rm ISCO}$ significantly affects the values obtained for $R_{\rm J}$, $\dot{M}_{\rm in}$ and $m_{\rm S}$. Also in this case, $R_{\rm J}$ decreases through the three phases, but starts from almost 44 R$_{\rm G}$ to then decrease to $\sim$20-30 R$_{\rm G}$ and finally to $\sim$15-20 R$_{\rm G}$ in phase 2 and 3 respectively. The trend is significant, since the discrepancy between these best-fit values is larger than their errors. A different trend is observed in $\dot{M}_{\rm in}$, which remained very stable throughout the eight Epochs analysed, with only a slight increase in phase 3. The stability in $\dot{M}_{\rm in}$ is compensated by $m_{\rm S}$ evolving blatantly, going from $m_{\rm S} \sim 1.1$ in phase 1 to $m_{\rm S} \sim 0.75$ and finally rising again in phase 3 up to 1.5. If we impose the same $m_{\rm S}$ for all the epochs, in conformity with the analysis performed on GX 339--4 \citep{Marcel2018a}, the best fits found by \textsc{Xspec} are strongly reflection-dominated. This happens because for $m_{\rm S}$ values beyond 1, the JED model is characterized by a rather high electron temperature. The subsequently high energy cutoff is not compatible with the spectral curvature well-traced by the BAT data in our spectra, which indicate a cut-off at about 100 keV. The only way for the \textsc{JED-SAD} model to reproduce such spectral curvature would be to decrease $R_{\rm J}$ and simultaneously increase the temperature of the inner edge of the SAD. However, a hotter SAD should also be visible in the XRT band, which does not show any sign of such strong disk component. Since the \textsc{JED-SAD} is not suitable, the best solution to describe the broadband spectrum consists in tuning up the inner (highly-ionized) reflection component. Such reflection-dominated spectra are challenging to explain, especially in the framework of the proposed JED-SAD geometry. Similarly high reflection fractions have been explained in the past with, e.g., a large covering fraction of the disk due to the presence of clouds of colder matter \citep{Malzac2002}, with a compact corona close to the BH emitting anisotropically due to light bending \citep{Miniutti2003} or due to a mildly-relativistic motion of the corona towards the disk \citep{Beloborodov1999}. Unfortunately all of these scenarios are hard to reconcile with the proposed JED-SAD geometry. We conclude that with $R_{\rm ISCO}=2$ R$_{\rm G}$ the role played by $\dot{M}_{\rm in}$ in driving the evolution of the accretion flow towards intermediate states is somehow replaced by an evolution in $m_{\rm S}$.

\subsection{Exploring the effect of the other JED-SAD parameters}\label{ss:distance}

In order to reduce the number of degrees of freedom, so far we kept some parameters fixed in both Models 1 and 2. In the following we explore in more detail their impact on the results obtained. \\
As shown by \cite{Marcel2018a}, fig. 10, $p$ (which was labeled $\xi$ by these authors) has little to no effect to the JED-SAD spectral shape. Similarly, the choice of fixing $\mu$ to 0.5 should have a negligible impact, especially since the high energy cut-off is basically insensitive to this parameter \citep[see fig. 6, ][]{Marcel2018a}. On the contrary, the jet power $b$ is expected to influence the JED shape at both hard and soft X-rays. Broadly speaking, by increasing $b$ more power is funneled in the jets, making the underlying JED rather cold, while with a lower $b$, the JED results hotter. This effect would shift the position of the high energy cut-off and therefore act as $m_{\rm S}$. However, $b$ is a function of the aspect ratio $\epsilon$, which for a typical JED should be about 0.2-0.3. As shown by \cite{Petrucci2010}, Appendix A, we need $b$ values around 0.3-0.5 to obtain such aspect ratios, as higher values would produce discs geometrically too thin to be consistent with the expected JED geometry. We are therefore not exploring the effect of varying $b$ any further. \\
Potential biases could be ascribed to the dilution factor $\omega$ and to the parameter $K_{\rm JEDSAD}$, which is solely related to the distance. We recall $\omega$ being a parameter regulating the amount of SAD photons penetrating the JED and thereby cooling it. In the current model, it ranges between 0 (the unlikely case where no photons from the SAD are intercepted by the JED) and 0.3 (when up to 30\% photons can be intercepted). In order to check the effect of $\omega$, we analyzed again Epoch 5 keeping $\omega$ fixed first to 0 and then to 0.3. As expected, when we put $\omega$ to zero, the cooling effect played by the SAD is replaced by lowering $R_{\rm J}$, which attains a value of 27-31 R$_{\rm G}$, while also $\dot{M}_{\rm in}$ and $m_{\rm S}$ drop down to values of about 1.1 $\dot{M}_{\rm Edd}$ and 1.0 respectively. Increasing $\omega$ to 0.3 only moderately impacts the outcomes of the fits, i.e. with $R_{\rm J}$ and $m_{\rm S}$ compatible with the results obtained with $\omega=0.2$ and a 15\% increase in $\dot{M}_{\rm in}$. The distance of \mysou\ is well constrained between 2.7 and 3.3 kpc \citep{Atri2020}. In the previous sections, we froze $\omega$ to 0.2 and $K_{\rm JEDSAD}$ to 11.11, corresponding to a distance of 3 kpc. We then computed again the best fits for Epoch 5 freezing $K_{\rm JEDSAD}$ first to 13.7 (2.7 kpc), and then to 9.2 (3.3 kpc). Analogously to $\omega$, variations of $K_{\rm JEDSAD}$ do not affect the overall physical scenario which was individuated in this phase phase with $K_{\rm JEDSAD}=11.11$, but they result in a slight fluctuation of the best-fit values of $\dot{M}_{\rm in}$, $R_{\rm J}$. For example, $R_{\rm J}$ goes from $\sim$41 R$_{\rm G}$ at $d$=2.7 kpc to $\sim$48 R$_{\rm G}$ at $d$=3.3 kpc, i.e. a range comparable with the estimated errors (about 2-5 R$_{\rm G}$). On the other hand, $\dot{M}_{\rm in}$ oscillates between $\sim$1.58 and $\sim$1.70 $\dot{M}_{\rm Edd}$ (for the lower and higher edge of the distance range respectively), which is comparably higher than the range of variability associated to the errors. These results point out that some caution must be taken when considering the best-fit values obtained for $R_{\rm J}$ and $\dot{M}_{\rm in}$ as face values.

\section{Discussion}\label{sec:discmaxi}
   
In the previous section we applied the physical model \textsc{JED-SAD} to fit the XRT+{\it NICER}+{\it NuSTAR}+BAT observations of the source MAXI J1820+070 in hard and hard/intermediate state. We included either one (Model 1) or two (Model 2) reflection components. In the latest, more complex model, the two reflection spectra originate from different but neighboring regions of the SAD disk. In all the Epochs analyzed, with the exception of Epoch 1 (the only one in full hard state), we found that only a 2-reflection solution gives acceptable fits. In the following, we will discuss the main results obtained in this work and their interpretation. \\
To fit our spectra we considered both a moderately ($R_{\rm ISCO}$=4 R$_{\rm G}$) and rapidly ($R_{\rm ISCO}$=2 R$_{\rm G}$) spinning BH. We note that in the case of a rapidly spinning BH, the evolution of the system is mainly driven by variations in the sonic Mach number $m_{\rm S}$. This solution is different with respect to the modelling of GX 339--4, in which $m_{\rm S}$ was fixed to 1.5 and the interplay between $R_{\rm J}$ and $\dot{M}_{\rm in}$ was sufficient to describe the data \citep{Marcel2019}. We observed that this trend is statistically significant, since imposing constant values of $m_{\rm S}$ results in unphysical reflection-dominated scenario.
A change in $m_{\rm S}$ arises from a change in the torque due to the jet (through the toroidal magnetic field). We see no clear reason for the jet torque to change on long (> days) time scale, even if it is possible in principle. Although we discard here this possibility, one should nevertheless keep it in mind and look for traces of any correlation between these rapid accretion (increase in $m_{\rm S}$) events and any jet signature. \\
On the contrary, in the fits performed keeping $R_{\rm ISCO}$ equal to 4 R$_{\rm G}$, the $m_{\rm S}$ values are (almost) constant and the interplay between $R_{\rm J}$ and $\dot{M}_{\rm in}$ plays again the main role. We therefore consider the fits performed in the moderately spinning BH scenario more physically reliable. This evidence is in line with the estimation of the spin value obtained by e.g. \cite{Fabian2020} and \cite{Guan2020}. Nevertheless, even in the case $R_{\rm ISCO}$=4 R$_{\rm G}$, a (slight) increase in $m_{\rm S}$ is observed going from phase 2 to phase 3, i.e. from 1.25 to 1.50. Imposing $m_{\rm S}$ to 1.5 in Epoch 6, in continuity with Epoch 7, results in a significantly worst fit where the reflection dominates over the direct JED emission. While this is slightly different with respect to the results obtained by \cite{Marcel2019, Marcel2020}, it is noteworthy that these authors did not have data above 40 keV and, most importantly, they did not directly perform spectral fits on the data. In future applications of the JED-SAD model, the inclusion of high energy \nustar\ and especially BAT data might therefore be necessary to highlight the role, if any, of $m_{\rm S}$ in shaping the spectra of BHTs in hard state and, particularly, their high energy cutoff.

\subsection{The geometry of the accretion flow}
The geometry of the accretion flow in the hard state of \mysou\ has been inspiring a vivacious debate since its discovery. Several authors have proposed that, when in hard state, the hot corona of \mysou\ contracts while the accretion disk remains stable at the ISCO: this scenario was proposed first by \cite{Kara2019} and recently supported by \cite{Wang2021}, based on \nicer\ X-ray spectral and timing analysis. These results were also corroborated by modeling the reflection component in \nustar\ spectra, adopting two lamppost coronae models \citep{Buisson2019,You2021}. However, \citet{Zdziarski2021} proposed that the truncated disk scenario
can still explain the hard and hard/intermediate state of \mysou. According to their spectral analysis performed with {\it NuSTAR} data on our Epochs 1,2,3 and 4, the disk never extends down to the ISCO. A scenario where the disk is not only truncated, but also the inner edge of the disk approaches the BH throughout the hard state is also suggested by the spectral evolution of the quasi-thermal component responsible for the thermal reverberation lags \citep{DeMarco2021} and the evolution of the characteristic variability frequencies observed in the iron line \citep{Axelsson2021}. It is noteworthy that a trend for an approaching truncated inner disk is also qualitatively suggested by Type-C QPOs \citep{Buisson2019}. \\
Regardless of the choice for $R_{\rm ISCO}$ and of the uncertainty in $K_{\rm JEDSAD}$, all the fits performed by us show that the accretion disk in MAXI J1820+070 is truncated in hard state and its inner edge moves inward during the transition to the intermediate state. For $R_{\rm ISCO}$=4 R$_{\rm G}$, we found $R_{\rm J}$ being about 50-60 R$_{\rm G}$ in phase 1, 35-45 R$_{\rm G}$ in phase 2 and 30-35 R$_{\rm G}$ in phase 3. A bit of caution is required in taking these measures as face values, since the exact estimates of the truncation radius are instead critically dependent on our choices for $R_{\rm ISCO}$ and $K_{\rm JEDSAD}$. Our spectral analysis was performed on a data set which includes the data set of \cite{Zdziarski2021} and reaches a similar conclusion on the geometry. However, our work includes four more observations and provides a coherent picture of the spectral evolution of the system, using a physical model. The calculated transition radii $R_{\rm J}$ are consistent with the lower limits on the truncation radius estimated by \cite{DeMarco2021} by analysing the quasi-thermal spectral component arising from the disk due to X-ray irradiation from the corona (see, in particular, their fig. 9). \\
Finally, a multi-zone corona was invoked on the basis of both spectral \citep{Zdziarski2021} and temporal \citep{Dzielak2021} behaviour of the system in the same observations. We highlight that the JED model is also a multi-zone model, as it takes into account the single contribution for each "ring" of matter which composes the JED, each of them with variable temperature and optical depth. Therefore, our approach is in line with other works, where a single uniform plasma cloud was proven insufficient to adequately describe the spectrum of Black Hole X-ray binaries  \citep{Nowak2011,Basak2017}. Moreover a multi-zone model can naturally produce a spectral shape which is not exactly a power law. In the energy range of the iron line, this can have a direct impact on the broadness of the line  and could explain the absence of extreme red wings that would require an accretion disk down to the ISCO. 

\subsection{On the nature of the second reflection component} 
\begin{figure*}
\centering
\includegraphics[width=\columnwidth]{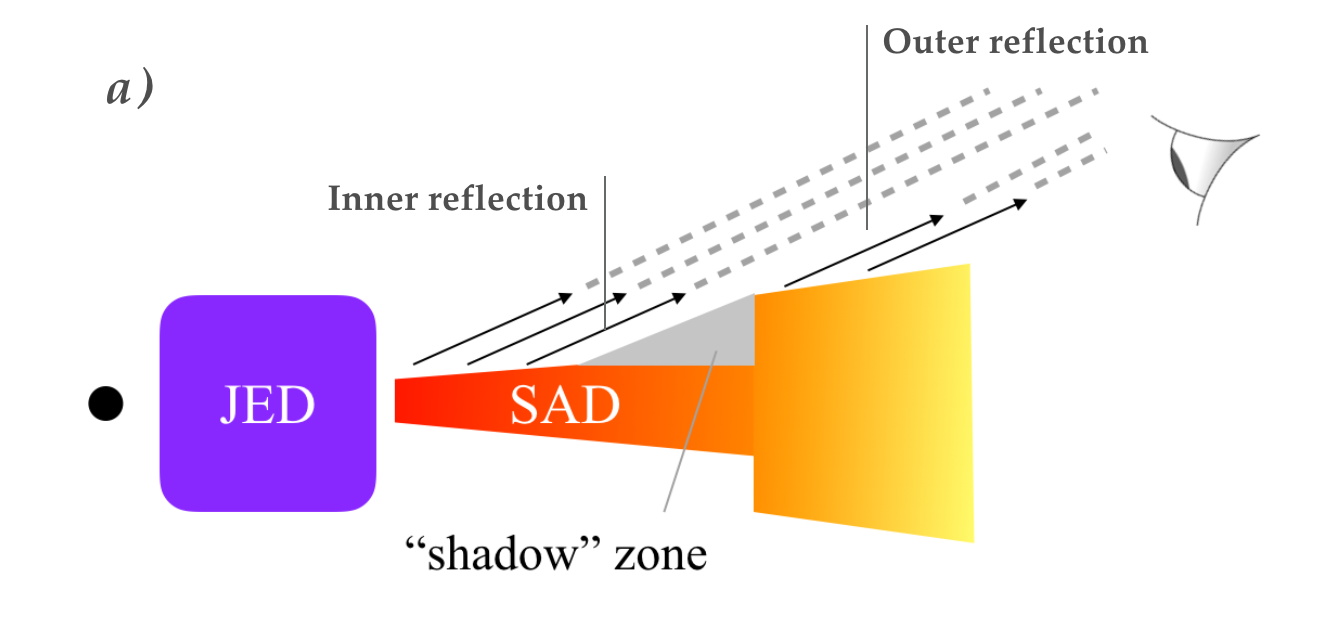}
\includegraphics[width=\columnwidth]{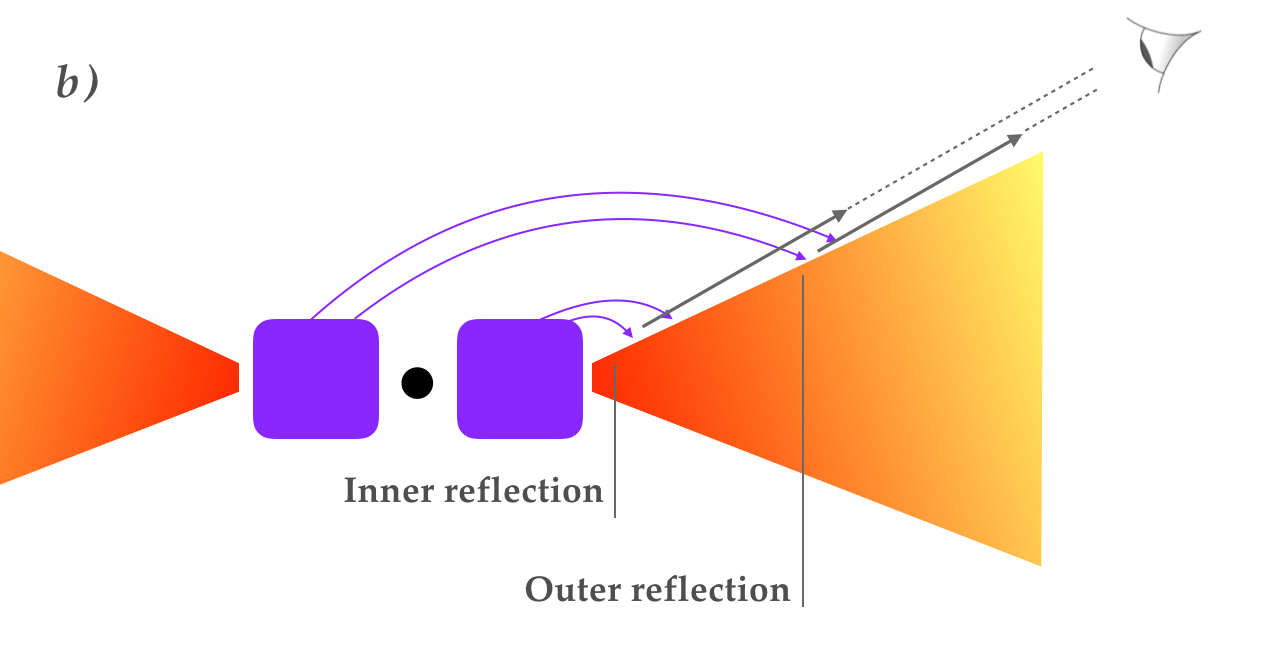}
\caption{Sketches representing two possible scenarios that might generate the double reflection component used to model X-rays data of MAXI J1820+070 in hard state. In particular, in panel {\it a} we show that a sudden jump in the scale height of the disk might shadow part of the reflection coming from the intermediate regions. In panel {\it b}, we show that the contribution from the on-the-other-side-of-the-BH JED could illuminate particularly the outer regions of the disk, leading to a boost in the outer reflection component.}
\label{fig:self-shielding}
\end{figure*}

In \cite{Kara2019} the line profile revealed by the {\it NICER} data was evidently evolving across the hard and hard/intermediate state, with a shrinking narrow core and a constant  broad, blurred base. This has been interpreted by invoking: (i) a disk reaching the ISCO in hard state, and (ii) a contracting lamppost corona, which produces two reflection components, one stable (the broad inner one) and one variable (the narrow outer one). In this geometry, the top of this corona would  mainly illuminate a distant region of the disk, which produces the narrow line, contrarily to the bottom of the corona, which would instead illuminate an inner region of the disk, where a strongly blurred line arises. Using two reflection components was also required in the spectral analyses by  \citealt[][]{Buisson2019,Chakraborty2020,Zdziarski2021}. In our modelling, the broad component in the iron line is also stable and changes very slightly from Epoch 2 to 7 (see Fig. \ref{fig:reflection_profiles}). However, the disk does not reach the ISCO and the broadening of the line profile is possibly due to the combination of two effects. Firstly, a significant role is played by the underlying continuum. Indeed, previous papers reporting  a highly ionized disc reaching close to the ISCO in the hard state consider a single Comptonization continuum, while a multi-zone Comptonization, as considered in the JED-SAD model, is more physically motivated. Furthermore, as a consequence of the high ionization parameter estimated for the plasma in the inner disk ($\log{\xi_1} \sim 3.7-3.8$), photons in the iron line tend to suffer multiple Compton scatterings, leading to a further broadening of the line profile \citep{Matt1996}.
In the JED-SAD configuration the corona does not have a lamppost geometry and an explanation analogous to \cite{Kara2019} can not hold. An alternative possibility could be the inadequacy of our reflection models in dealing with real-life reflection spectra. Indeed, we note that the reflection spectrum observed in accretion disks results always from the {\it continuum} of reflection spectra emitted at different radii. Using only one reflection component could just be an oversimplified approximation in this case. Nevertheless, given the high inclination of the system, some effects could also create an imbalance in the contributions from the different parts of the disk, i.e. lowering the contribution from the intermediate regions or increasing that from the outer regions. We suggest two mechanisms potentially responsible for such imbalance, sketched in Fig. \ref{fig:self-shielding}. 

\begin{figure}
\centering
\includegraphics[width=\columnwidth]{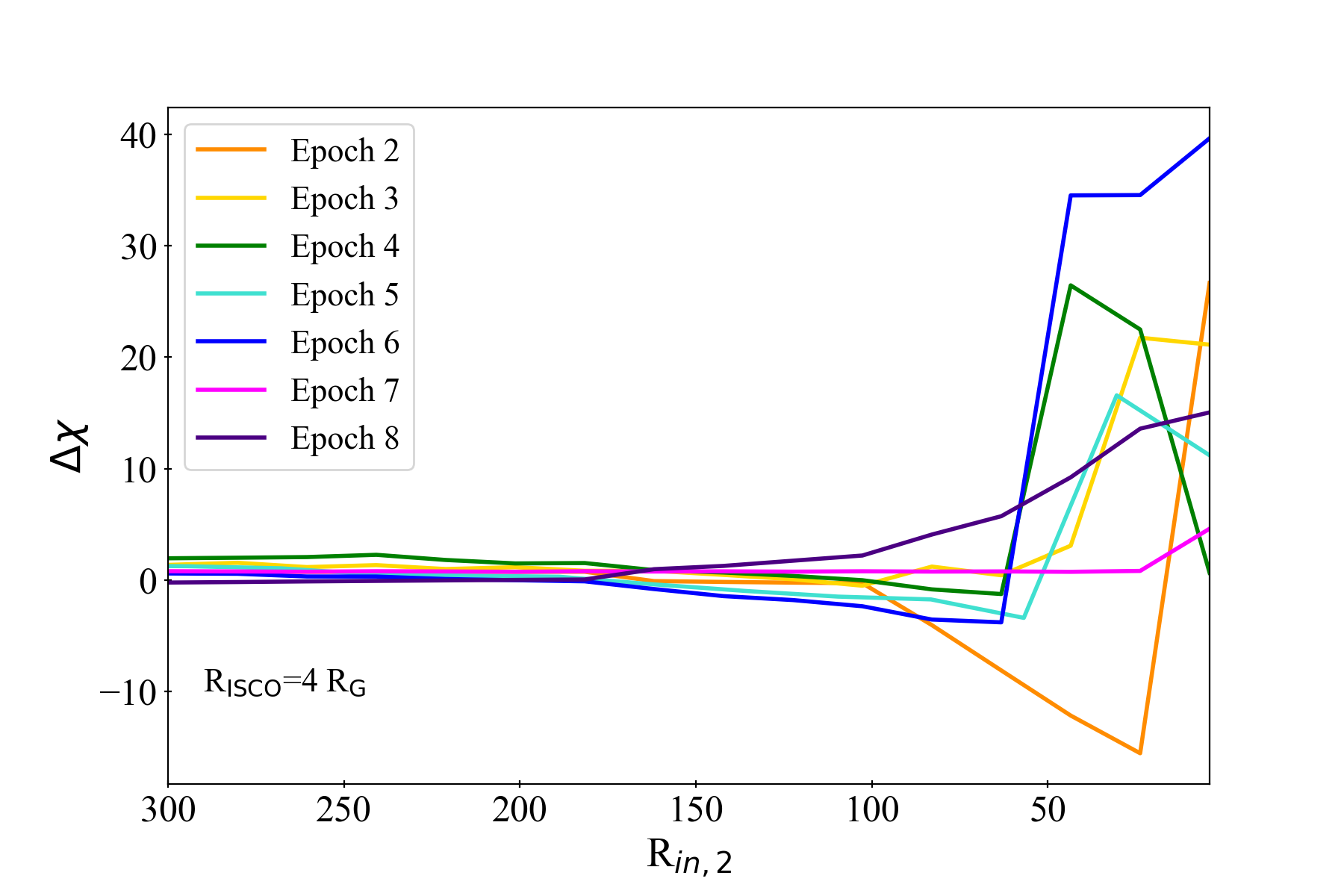}
\caption{Contour plots for the parameter $R_{\rm in,2}$ when left free and not tied to $R_{\rm J}$.}
\label{fig:contour}
\end{figure}

One possibility could be that the disk, since viewed at such high inclination, might create a "shadow" zone where part of the reflection incoming between $R_{\rm in,1}$ and $R_{\rm in,2}$ is obscured (see  Fig. \ref{fig:self-shielding}, panel {\it a}). As highlighted by the Figure, a "jump" in the scale height of the disk, located sufficiently far away from its inner edge, could in principle generate such self-shielding effect. A disk "flared" in the outer region was also put forward by \cite{Zdziarski2021} and \cite{Axelsson2021} in order to explain the presence of the outer reflection component and the variability associated to the iron line. However, it is not obvious whether a sudden puffing up of the disk could happen or not in a classical Shakura-Sunyaev disk, like the SAD model. A jump could arise from a transition in the mechanism contributing mainly to the opacity in the disk. In the outer regions of a Shakura-Sunyaev disk, the opacity is expected to be mainly due to free-free absorption, while Thomson scattering dominates the opacity in the inner regions of the disk \citep[see equations (2.16) and (2.19) of ][]{Shakura1973}. The geometrical thickness of the disk scales with the radius $R$ following a (slightly) different relation, i.e. it is proportional to $R^{-21/20}$ in the inner, Thomson scattering dominated, regions and to $R^{-9/8}$ in the outer regions. At the boundary between these two zones, i.e. at a radius $R_{\rm jump}$, a jump is expected. We calculated $R_{\rm jump}$ according to our results for $\dot{M}_{\rm in}$ and assuming the same value for $\alpha$ in both equations. We found $R_{\rm jump}$ between 3$\times$10$^4$ R$_{\rm G}$ and 7$\times$10$^4$ R$_{\rm G}$, in all the epochs considered. In our fits, the location of the outer reflector $R_{\rm in,2}$ was kept frozen to 300 R$_{\rm G}$. When the parameter is thawed, it is completely unconstrained, as shown in Fig. \ref{fig:contour}). Notwithstanding the uncertainty on $R_{\rm in,2}$, a boundary at beyond 10$^4$ R$_{\rm G}$ is likely located too far away to produce an extra-reflection component. However, a crucial role might be played by irradiation from the inner regions of the disk to the outer regions, an ingredient which is neglected in Shakura-Sunyaev disks. Beyond some boundary radius, we expect that the upper layers of the disk should be heated up by the impinging photons coming from the inner regions, leading to evaporation and to a larger effective scale height.

Such effect would be efficient only where thermal winds are launched and unfortunately this region is also located beyond 10$^5$-10$^6$ R$_{\rm G}$, i.e. where the sound speed of the plasma overcomes the Keplerian speed \citep[see, e.g. ][]{Higginbottom2017}. This argument would exclude that the optical winds detected in \mysou\ in hard state \citep{MunozDarias2019, SanchezSierras2020} could play the role of shadowing reflection from intermediate regions. However, the presence of a large scale vertical magnetic field make real-life accretion disks more layered and "puffed up" \citep{Zhu2018,JacqueminIde2021}. It is therefore possible that the boundary radius at which the disk is geometrically thickened by irradiation may be located even closer and reasonably impacts the reflection spectrum observed from the disk at high inclinations. In order to confirm this, irradiation must be properly taken into account in models of magnetized disks around BHs\footnote{As done e.g. in \cite{Zhu2021} for Young Stellar Objects.}. However, this is way beyond the scope of the present paper. 

An alternative scenario is sketched in Figure \ref{fig:self-shielding}, panel b. In principle, the radiation from the JEDs on the two different sides of the BH with respect to our line of sight could illuminate two differently located regions of the SAD, originating the two reflection components. Indeed, photons coming from the other-side-of-the-BH JED may not be able to impinge on the SAD before $R_{\rm in,2}$, producing an excess in the reflection component coming from this radius. The picture presented here is likely oversimplified, as it may require a finely tuned light bending of the radiation coming from the other side of the BH and it does not take into account any possible occultation from the jet, expected to be optically thin. \\
In addition, it is unclear how the evolution, mainly in ionization, of the outer reflection component fits within the proposed scenarios (see Fig. \ref{fig:reflection_profiles}). Such trend could be explained with an approaching of $R_{\rm in,2}$, which was instead kept frozen to the assumed value of 300 $R_{\rm G}$. Indeed, as shown in Fig. \ref{fig:contour}, $R_{\rm in,2}$ remains essentially unconstrained by the fit, so that an approaching trend may be hidden here. If we consider the irradiation from the inner disk as responsible of the puffing up of the outer disk and the subsequent self-shielding effect, the fading in X-ray luminosity of the system during Epochs 2 to 8 could be somehow correlated to the observed variability in the outer reflection. However, this can not be confirmed without a proper MHD modelling of magnetized disks that takes into account how irradiation impacts the scale height of disks.

\section{Conclusions}

In this paper, we reported on the spectral study of the BH transient MAXI J1820+070 in hard state exploiting the JED-SAD accretion-ejection paradigm. This is the second object the model has been applied to and the first time that Compton reflection is taken into account in detail. We investigated the spectral behaviour of the system in 8 epochs, spanned over roughly 100 days, during which the system was always in hard and hard/intermediate state. Due to the uncertainty on the spin of the BH, we considered two possible values for $R_{\rm ISCO}$, i.e. 4 R$_{\rm G}$ (a$_*$=0.55) and 2 R$_{\rm G}$ (a$_*$=0.95), respectively. In the first case, only two parameters, $R_{\rm J}$ and $\dot{M}_{\rm in}$, drive the largest changes in the spectral modelling: $R_{\rm J}$ decreases throughout the period, while $\dot{M}_{\rm in}$ first increases, i.e. during the rise, and then decreases, i.e. during the plateau and the decline \citep[see ][ for the definition of these phases]{DeMarco2021}. For a rapidly rotating BH, $\dot{M}_{\rm in}$ remains quite constant, while a decrease in the sonic Mach number $m_{\rm S}$ is observed and seems to drive the overall evolution of the accretion flow. This scenario is unlikely and suggests that the BH in \mysou\ is rather moderately spinning, as also found by \cite{Fabian2020} and \cite{Guan2020}. The geometry of the system consists in a truncated disk, i.e. $R_{\rm J}$ never goes below 15 R$_{\rm G}$, with the inner radius decreasing during the monitored period, in agreement with e.g. \cite{Zdziarski2021} and \cite{DeMarco2021}. In order to successfully describe the spectra, two reflection components have to be taken into account for  all the epochs considered (except for the first one): one component is highly ionized and originating from the edge of the SAD, while the other is less ionized and presumably originates from an outer region of the disk. While the inner reflection component is stable, the outer reflection component evolves in a way that is compatible with a scenario where the region responsible for such a component is approaching the BH. A self-shielding effect, due to both the high viewing angle and a flared outer disk, might offer a viable explanation for such a phenomenon. \\ It is worth noticing that the exact $\dot{M}_{\rm in}$-$R_{\rm J}$ values estimated here depend on the choice of $R_{\rm ISCO}$, the uncertainty on the distance of the system and some simplifying assumptions on the dilution factor $\omega$ and the jet power fraction $b$. However, we argued that these assumption do not affect the main conclusions provided by this work, i.e. the truncated disk geometry and the double reflection component, but only provide additional uncertainty on the best-fit values presented here. Further investigations are necessary to obtain more reliable and precise measurements of the truncation radius and the inner mass-accretion rate. One possibility would be to check if the obtained $R_{\rm J}$ and $\dot{M}_{\rm in}$ values can be used to fit the radio observations of this system, as successfully done by \cite{Marcel2019} for GX 339--4. Results discussing the jet behavior and brightness using the JED-SAD model for \mysou will be presented in a forthcoming publication.

\begin{acknowledgements}
AM, MDS, AS, SM, ADA, TDS and TDR acknowledge a financial contribution from the agreement ASI-INAF n.2017-14-H.0 and from the INAF mainstream grant (PI: T. Belloni, A. De Rosa). AM and TDS acknowledge financial contribution from the HERMES project financed by the Italian Space Agency (ASI) Agreement n. 2016/13 U.O. POP, SB and JF acknowledges financial support for the CNES space national agency and the CNRS PNHE. We thank R. La Placa for fruitful discussion.
\end{acknowledgements}

\bibliographystyle{aa}
\bibliography{biblio}

\end{document}